\documentclass{article}

\usepackage{arxiv}
\usepackage{longtable}
\usepackage{lscape}
\usepackage[utf8]{inputenc}  
\usepackage[T1]{fontenc}     
\usepackage{hyperref}        
\usepackage{url}             
\usepackage{booktabs}        
\usepackage{amsfonts}        
\usepackage{nicefrac}        
\usepackage{microtype}       
\usepackage{lipsum}
\usepackage{graphicx}
\graphicspath{ {./images/} }
\usepackage{booktabs}
\usepackage{float}
\usepackage{graphicx}
\usepackage{lscape}
\usepackage{array, makecell}

\title{Automation from the Worker's Perspective: \\
How can new technologies make jobs better?}

\author{
Ben Armstrong \\
 Industrial Performance Center \\
 Massachusetts Institute of Technology \\
 Cambridge, MA 02139 \\
 \texttt{armst@mit.edu}
\And
 Valerie K. Chen \\
  Electrical Engineering and Computer Science\\
  Massachusetts Institute of Technology\\
  Cambridge, MA 02139 \\
\And
 Alex Cuellar \\
  Aeronautics and Astronautics\\
  Massachusetts Institute of Technology \\
  Cambridge, MA 02139\\
\AND
  Alexandra Forsey-Smerek \\
  Aeronautics and Astronautics\\
  Massachusetts Institute of Technology \\
  Cambridge, MA 02139\\
\And
  Julie Shah \\
  Aeronautics and Astronautics\\
  Massachusetts Institute of Technology \\
  Cambridge, MA 02139\\
}

\begin{document}
\maketitle
\begin{abstract}
Common narratives about automation often pit new technologies against workers. The introduction of advanced machine tools, industrial robots, and AI have all been met with concern that technological progress will mean fewer jobs. However, workers themselves offer a more optimistic, nuanced perspective. Drawing on a far-reaching 2024 survey of more than 9,000 workers across nine countries, this paper finds that more workers report potential benefits from new technologies like robots and AI for their safety and comfort at work, their pay, and their autonomy on the job than report potential costs. Workers with jobs that ask them to solve complex problems, workers who feel valued by their employers, and workers who are motivated to move up in their careers are all more likely to see new technologies as beneficial. In contrast to assumptions in previous research, more formal education is in some cases associated with more negative attitudes toward automation and its impact on work. In an experimental setting, the prospect of financial incentives for workers improve their perceptions of automation technologies, whereas the prospect of increased input about how new technologies are used does not have a significant effect on workers' attitudes toward automation.\footnote{The survey on which this research is based was conducted by IPSOS and sponsored by Amazon. The authors designed the survey and conducted the research independently. The authors are grateful to the Siegel Family Endowment for financial support. The authors have also received support through the MIT - Amazon Science Hub to advance research on industrial automation.}
\end{abstract}

\section{Introduction}
Sam Sayer worked in factories for years before he ever saw a robot. Now, in a red beard and zombie t-shirt, he’s teaching his colleagues on the factory floor how to use them \cite{mit_industrial_performance_center_factories_2024}. Sayer is one of a growing group of employees at a New Hampshire cutting tool manufacturer who have become champions of the automation transforming their jobs. In recent years, Sayer’s employer has scaled advanced technologies across the company. Robots and advanced machines are distributed throughout their factories, working alongside employees to make them faster and freeing workers up to solve problems that arise on the floor.

The technological transformation has coincided with new job opportunities for the workforce. In addition to operator roles, workers can move into process support and manufacturing process roles, requiring more skills and earning them higher pay. Sayer’s employer relies on new technologies to compete in the global economy and grow their enterprise. As a result, they also rely on workers like Sayer to learn about the new technologies and embrace them as a way of making their jobs and careers better. But this is not always the case.

At an aerospace manufacturer in Ohio, there was a similar ambition to use automation to improve work and make the company more competitive. The company hired an integrator to train the robot and even gave it a name to make the new “teammate” more palatable to workers on the factory floor. But the workers in Ohio did not embrace the technology. Some worried it would be a threat to the core of their roles as machine operators. Others did not find it particularly useful. It stopped often, and they could load and unload parts faster than the machine.

A series of studies across industries have shown that when firms adopt new technologies, they become more productive, more competitive and hire \textit{more workers} \cite{dinlersoz_automation_2018,dixon_robot_2021,hirvonen_new_2022}. However, the impact of new technologies on the quality of work is less clear. Some workers embrace new technologies as beneficial for their jobs, whereas other workers worry about the impact of automation or resist it.

With the rapid development of new technologies that promise to reshape jobs –- from robotic systems to AI tools –- we are interested in how new technologies can make jobs better.

Understanding how to harness the benefits of new technologies is important for individuals, organizations, and society. At the \textbf{individual level}, work is where a great many people spend a majority of their waking hours. It can lead to injury, lead to mental health challenges, and be a source of fulfillment and personal growth. Understanding how to improve the quality of jobs can make individuals healthier and happier.

The relationship between technology and job quality also matters at an \textbf{organizational level}. Recruiting and retaining skilled workers is key for employers across sectors. And for organizations motivated to adopt new technologies and improve, a core challenge is finding workers who will champion or embrace new technologies that can make firms more competitive. Understanding what practices help employers compete, while also making their organizations good places to work, enables what we refer to as "positive-sum automation," or technological change that benefits employers and their employees -- and contributes to productivity as well as flexibility.

At a \textbf{societal level}, the public dialogue about technology and work is often bipolar. Some arguments emphasize the potential for technology to displace work or make it obsolete. Others suggest technologies will eliminate drudgery, allowing workers to be more creative and only focus on the most engaging parts of their jobs. Both perspectives are frequently presented as an inevitability, rather than a contingent outcome that will depend on individual and organizational choices, as well as the design of new technologies.

This study aims to identify the factors affecting worker attitudes toward automation based on the perspectives of workers themselves --- and their experiences with new technologies on the job. It introduces new data from a multinational survey of more than 9,000 workers to highlight the worker’s perspective. The survey results enable us to answer a multidisciplinary set of questions with the potential to influence employment strategy, training, and technology design.

The survey finds that workers are generally more positive than negative about the impact of automation and new technologies on various aspects of their work, including their safety and comfort, autonomy, pay, upward mobility, and job security. Much of this paper focuses on why some workers see more benefits in new technologies than others. We identify five factors that predict whether workers see new technologies and automation as beneficial for their jobs.

\paragraph{Cross-national variation.} \textit{Perceptions of automation are broadly positive across countries, although Americans are most pessimistic}. The pessimistic attitude toward new technology among Americans is surprising given historically optimistic attitudes toward innovation and technological change among U.S. companies. In countries with strong social safety nets such as France, Germany, Italy, and Spain, the perceived impact of automation on wages and job security was most positive, whereas in the most liberal market economies – Australia, the United Kingdom, and the United States – the perceived impact of automation on job security and pay was neutral or negative.

\paragraph{Job tasks.} \textit{Workers performing jobs that require complex problem solving or new ideas tend to be more positive about automation}. Although economists have long identified workers in routine jobs as most vulnerable to automation, workers’ level of routine tasks does not seem to be the strongest predictor of how they think about the impact of automation on their work. Instead, workers report doing a variety of tasks as part of their jobs with complex problem-solving tasks as a strong predictor of workers seeing new technologies as beneficial.

\paragraph{Education and race.} \textit{Although new technologies are often presented as “skill-biased,” benefiting more educated workers, perceptions of automation sometimes follow the opposite pattern}. Workers with less formal education are often more optimistic about the impact of automation on their work. Also, Black and Hispanic workers in the United States are far more optimistic about the impact of automation on their jobs than workers of other racial backgrounds.

\paragraph{Employers matter.} \textit{Workers who feel like their employer values them and is invested in their safety are more optimistic about the impact of automation at work}. Workers’ job satisfaction and level of trust are also highly predictive of their attitudes toward automation. There is some indication that workers with employers that have previously invested in technology adoption are more optimistic about automation.

\paragraph{Attitudes toward work.} \textit{Workers who are motivated to learn new skills and grow in their careers are more likely to see automation as a positive force at work}. A minority of workers report that upward mobility is a key factor to making a good job, but a high share of workers report that employers do not invest sufficiently in learning and growth opportunities. Those workers for whom upward mobility is important are more optimistic about new technologies and automation than their peers. 

The remainder of the paper is divided into six sections. It reviews the motivation for the survey (Section II), the questions the survey posed (Section III), the hypotheses it sought to test (Section IV), descriptive statistics about workers’ attitudes (Section V), as well as key findings (Section VI). It will also propose potential lessons for policymakers, employers, and the public (Section VII).

\section{A Bottom-up Perspective on Automation and Work}

Workers’ relationship to technology is a key topic for multiple social science and engineering disciplines, each with different methodological tools and areas of emphasis. Some scholars have focused on the broad economic implications of new technologies for workers. Others have focused on how technologies can be designed to make particular tasks more enjoyable –- or less painful –- for individual workers.

Across a vast body of scholarship in economics, sociology, and political science, there have been four categories of findings about how technologies can make work better.

First, labor economists have concluded that technologies complement some workers and compete with other workers depending on the types of tasks that they perform at work. For example, some technologies like robots are considered “routine-biased,” which means that they compete with workers who perform routine tasks, but can complement workers who perform more cognitive or creative tasks at work, particularly when they are variable (non-routine) \cite{acemoglu_automation_2019, autor_growth_2013,acemoglu_skills_2011,goldin_race_2008}.

From this perspective, technologies create winners and losers among workers, creating more or less opportunities depending on the tasks inherent in certain jobs. Although the literature on skill-biased and routine-biased technological change has helped explain broad changes in the labor market, there are important caveats and critiques of this approach \cite{schmitt_dont_2013}.

Second, research in sociology and management has argued that technologies can provide better outcomes for workers when workers themselves have a voice in shaping how the technologies are adopted and implemented in the workplace. In these studies, a deliberative process in which unions or other worker representatives can provide input on how the technology is used can provide benefits for workers --– leading to technology adoption in ways that they prefer --– as well as firms, since worker championing new technologies could translate into higher productivity improvements \cite{litwin_technological_2011,kellogg_ai_2022,kochan_worker_2019}. Examples of this approach were on display in the SAG-AFTRA union’s negotiation over how generative AI tools could be introduced to replace or augment tasks that actors and writers had previously performed \cite{litwin_opinion_2023}.

Third, perceptions of technology may differ by location with workers in some countries proving more open to adopting new technologies than others. Research in comparative political economy provides a framework for thinking about why some workers might be more open to technological change than others \cite{hall_varieties_2001,thelen_regulating_2018}. Workers in countries with a stronger social safety net –- which includes stronger job security, as well as more generous support for those out of a job –- might be less concerned about the impact of new technologies on their job security, for example. By contrast, workers in countries with extensive labor regulations and good work conditions may be less optimistic about the ability for new technologies to improve their quality of work.

Fourth, adopting new technologies can improve productivity for those organizations and workers adopting them, which can translate into higher wages. Economists have documented a longstanding association between technology adoption, productivity and wages at the firm level \cite{dinlersoz_automation_2018,dixon_robot_2021,acemoglu_competing_2020,hirvonen_new_2022}. However, since the 1980s, there has been concern about a gap between economy-wide productivity and wages for workers without a college degree. Nonetheless, recent research has claimed that flattening wages for workers without a college degree arises from other factors –- and there is still a discernible contribution of productivity growth to workers’ wages \cite{stansbury_productivity_2017}.

Robotics and human factors engineering research has approached the relationship between new technologies and job quality by zooming in on particular technologies and tasks. Through experiments with individual users of technologies performing different activities, this research has identified patterns of how humans can use technological tools most effectively.

The scholarship highlights at least four areas that are important for determining how individuals relate to advanced technologies and automation. First, human characteristics and attitudes such as trust can shape how willing individuals are to use or resist a technology \cite{dzindolet2003-role, schaefer_meta-analysis_2016}. Second, the work environment and particular tasks workers perform could affect whether the technology is perceived as useful \cite{gombolay_decision-making_2015}. Third, workers with skills that enable them to make better use of technology (e.g. coding, dexterity) will operate the technology differently. And finally, the design of the technology can influence how it appeals to different groups \cite{manja2008-domestic, goetz2003-matching, zlotowski2020-one, sanneman_state_2021}. For example, designs can appeal to individuals with less familiarity with a technology or can be aimed to overcome attitudes or fears that might lead individuals to resist using a technology.

These bodies of literature have often succeeded in informing policy, organizational strategy, and technology design. However, while they aim to capture the impact of technology on workers’ jobs, these studies do not quite measure workers’ experiences with new technologies directly. In social science research, studies often impute what is good for workers based on wage data. And in engineering research, lab simulations or narrow surveys stand in to evaluate what workers prefer when it comes to new technologies.

A missing perspective in all this research is what workers themselves think about various new technologies under different conditions. The worker perspective can be illuminating for several reasons. For one, it can answer new questions that lab experiments and economic data cannot. For example, asking workers about what motivates them in their work can highlight whether workers motivated by pay might perceive new technology differently from workers motivated by comfort. These findings could lead to different approaches to technology implementation depending on the types of workers in the environment. 

Understanding the worker perspective is also an opportunity to test some of the ideas generated from previous economic and engineering studies. Economists have long proposed that automation and new technologies affects jobs differently depending on the types of tasks that those jobs are responsible for performing. Routine jobs, for example, are more vulnerable to technological change than jobs with more cognitive tasks. But different workers in the same job might report performing different tasks. By gathering task data and asking workers about the perceived impact of automation, the survey can test the relationship between tasks and the impact of automation from the worker perspective \footnote{Of course surveys have their drawbacks as opposed to alternative methods of data collection, such as observing what tasks workers pursue in their own environment. Workers might have different interpretations of what counts as routine, for example, making the data hard to compare. Nonetheless, these survey data are valuable as a reflection of what workers think about the task structure of their jobs.}. 

The results of the survey are meant to supplement the findings of previous work with new data from a different source. In addition to testing existing hypotheses, the survey is also positioned to generate new hypotheses that research in economics and engineering can test with alternative data sources.

\section{The Survey}

The method of capturing the worker’s perspective across a variety of work environments was a multi-country representative survey that asked respondents about the nature of their jobs, their attitudes toward work, their skills and tasks, as well as their perceptions of how positively or negatively technologies affect their work. The survey, conducted by polling group IPSOS, reached more than 9,000 workers across 9 countries (Australia, France, Germany, Italy, Japan, Poland, Spain, the United Kingdom, and the United States) in January 2024.

The survey is broad in scope, posing more than 200 questions divided roughly into five substantive categories (in addition to collecting demographic information).

The first captures workers’ work environments, asking about their occupation, industry, and pay, as well as the tasks they typically perform at work. The first section also asks workers about their relationship to their employer, including how their employer invests in various aspects of their job, including their pay, productivity, safety and comfort at work, learning and career development, and control over what you do and when you do it. 

To get a sense of the worker’s individual role, the survey asks workers about their individual tasks, and how much they enjoy them. The categories of tasks include collaborating with others, performing routine tasks, working with your hands, solving complex problems, and generating new ideas.  It asks workers about the technologies that they are exposed to at work, as well as the frequency that they use them within their jobs. The technologies range from robots to software automation tools, artificial intelligence tools, and generative AI. 

The second section of the survey also asks workers about how they interact with technology as part of their jobs. The responses include design technology for others to use, use technologies others have designed, oversee how others use technology, and troubleshoot technology to solve problems that arise during use.  There is also a series of questions about the information and skills that users need to operate technology effectively within their work.

The potential information workers will need includes how people typically use the technology tool, how the technology tool is supposed to perform when it is working well, how the technology works (what is going on under the hood), and how people need to interact with the technology for it to perform a specific task. Finally, workers are asked about the extent to which they are a technology leader or champion in their firm. The question asks the extent to which the person is a strong advocate for new technologies and the extent to which they are relied on by others within their firm for knowledge about the technology.

The third part identifies some of workers’ attitudes toward work and technology. It includes a module identifying workers’ inclinations to trust, including trust artificial intelligence. It also asks workers how various factors contribute to the quality of their job. The categories include pay, productivity, safety and comfort at work, learning and career development, and control over what the worker does and when they do it. Workers are asked to evaluate the extent to which each of these factors makes a good job. This section measures workers’ familiarity with various new technologies (independent of their use of those technologies at work). Building on past research, it also asks workers about their screen time and video game playing habits as potential predictors of their attitudes toward and experience with technology in general.

The fourth section of the survey includes a series of questions that ask workers about the perceived impact of various technologies –- in various conditions –- on aspects of their work (and the work of their co-workers). The features of their work we include are your productivity, your pay, the amount of control over what you do and when you do it at work, your ability to develop new skills and grow in your career, and your job security. This series of questions measuring how workers perceive the benefits versus the costs of automation and new technologies is the primary dependent variable for the study.

The fourth section also includes an embedded experiment. Respondents are asked to assess the impact of a hypothetical new technology, Alpha, on their jobs. The workers are randomly assigned to one of three groups. The first group is presented with a message board describing the potential impact of the technology and told that their employer will be giving them the technology to use. The second group is given the same message board, then told that they will be able to provide input on how the technology could be used best. The third group, again presented with the same message board, was told that they would be provided financial incentives –- a bonus –- if technology improved their productivity on the job. Workers in each of the three categories was then asked a series of questions about how they anticipated the technology would affect your productivity, your pay, the amount of control over what you do and when you do it at work, your ability to develop new skills and grow in your career, and your job security (see Appendix for treatment block details).

In the fifth and final section of the survey, workers saw multiple different technology designs: two humanoid robots, two robot arms, and two methods of conveyance (a conveyor belt and a mobile robot). In each of these categories, the images of technologies were paired such that one image was of a more ``playful''-looking design, and the other of a more ``machine-like'' design. They were then asked about their attitudes toward the different robot designs in various work environments. For example, how useful did they feel each of the two humanoid robots would be in a construction environment, as compared to a hospital environment? The survey module on technology design gave workers pictures of actual automation technologies and allows for comparison of worker perceptions among different versions of the same core technology (e.g. humanoid robot), as well as comparison between different technology categories (e.g. humanoid robot versus robot arm).

Survey participants are randomly assigned one of three automation categories: humanoid, robot arm, mobile technology. In each of these categories, the images of technologies were paired such that one image was of a more ``playful''-looking design, and the other of a more ``machine-like'' design: the humanoids shown were Pepper and Phoenix, robot arms were DOBOT and SZGH, and mobile technologies were an autonomous mobile robot (AMR) and conveyor belt, respectively. Participants first are shown the technologies and asked a series of questions on comfort around the technology and desire to use the technology. These responses allow for comparison of worker perceptions among different versions of the same core technology (e.g. humanoid robot), as well as comparison between different technology categories (e.g. humanoid robot versus robot arm). Then, participants are shown the technologies and images of restaurant, hospital, and construction settings, respectively representing a low-severity, mid-severity, and high-severity setting. They are asked a series of questions to compare the two technologies in each setting. As one image in each category is chosen to appear more "machine-like" than the other, the degree to which the responses prefer one technology over another for each setting can provide insights into the severity match of technology and task.

\begin{figure}
    \centering
    \includegraphics[width=.75\linewidth]{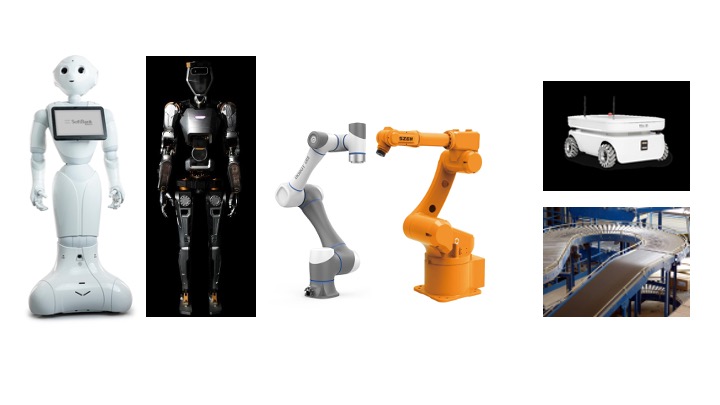}
    \caption{Automation Technologies in Survey\textsuperscript{a}}
  \small \textsuperscript{a} From left: Pepper, Phoenix, DOBOT, SZGH, TIAGo OMNI AMR (top-right), conveyor (bottom-right)
    \label{fig:enter-label}
\end{figure}

\section{Hypotheses and Methods}

The survey is designed primarily to understand how multiple factors –- work environment, employer relationship, engagement with technology, attitudes toward work and technology, and technology design (independent variables) –- are associated with a worker’s perception of the impact of automation and new technologies on various aspects of their work (outcome variables). In some cases, there are established hypotheses that we are testing. For example, previous studies have suggested that an individual’s level of trust, or the degree to which their job is routine, might alter the impact of technology on their work. 

In other cases, there is an intuition that the factor may be important, but the exact hypothesis is not clear. For example, it seems plausible that how a worker conceives of a “good job” may affect how they think about automation. However, it is not immediately clear whether workers motivated by pay would be more or less likely to think positively about the impact of automation compared to workers motivated by safety and comfort in a job. 

The hypotheses here are divided by the category of the survey. Tables  in the results section provides a summary of the hypotheses along with the results for each.

Each hypothesis also references the method by which it is tested. In many cases, the test is a multivariate regression model with the worker's perceived impact of automation as the dependent variable.\footnote{The multivariate regressions uses survey weights for each individual to ensure results are representative of each geography.} The most common dependent variable is a composite measure of worker's perceived impact of automation, which is the average of worker's perceived impact of automation across the five categories in the survey: safety and comfort, autonomy, pay, upward mobility, and job security. Some hypotheses may test the impact of a given factor (e.g. trust) on a different dependent variable, such as workers' perceived impact of automation on job security or pay alone. In these cases, the method will change to use a non-parametric test (if the dependent is a single likert scale variable) or a binomial logit model (if the dependent variable is a binary measure). The Appendix includes a glossary of variables used in the models.

\subsection{Work environment}
    
    Models of the relationship between technology and work often focus on the work environment itself: how much skill is required to do a particular job? What kinds of tasks are involved? Economists have associated jobs with less skill and more routine tasks with more vulnerability to automation. 
    
    As a result, we aim to test two related hypotheses from the worker’s perspective. Hypothesis 1A is that workers who are more educated will be more favorable toward the impact of automation technologies at work. Hypothesis 1B is that workers who perform certain combinations of tasks as part of their job will view automation more favorably. Research on routine-biased automation suggests that workers who perform fewer routine tasks are less vulnerable to automation and in this research would view the impact of automation on their work more favorably. By contrast, workers who perform more cognitive tasks --- which we define as "complex problem-solving" tasks in the survey --- will view the impact of automation more favorably. 
    
    In testing these hypotheses, the survey departs from previous research on tasks and automation. Labor economics studies have often modeled jobs as either routine or non-routine (cognitive), manual or non-manual \cite{acemoglu_skills_2011,autor_growth_2013}. The survey allows workers to identify the frequency with which they perform each set of tasks, allowing some workers to identify their jobs as both frequently routine and frequently cognitive, for example. The test of each aspect of Hypothesis 1B is whether a worker's frequency of routine or cognitive tasks predicts their perception of the impact of automation on their work.
    
    We also propose related hypotheses that are often taken as granted among engineers developing automation technologies. Hypothesis 1C is that workers will be more favorable toward technologies that help with frustrating, hard-to-do tasks. These are technologies that some might call “augmenting technologies.” Hypothesis 1D is that workers will be less favorable toward technologies that take over tasks that are core to their jobs, which some might call “automating technologies.” Although the ideal test for these hypotheses would be to measure the perceived impact of different technologies in a worker’s real-life work environment, we can use the survey to test proxies for these hypotheses understanding how workers perceive the impact of automation on their safety and comfort, as well as their autonomy (Hypothesis 1C) as well as on their job security and upward mobility (Hypothesis 1D). The test for these hypothesis relies on descriptive data (net favorability of each factor in the survey), rather than a discrete statistical test.
    
    There is also a potential relationship between workers and their employers that we seek to test. Hypothesis 1E is that if workers see their employer as invested in them, they will be more likely to look favorably at automation. In addition to a general measure of employer investment -– does the worker feel valued? –- the survey also asks workers the extent to which they find their employers invested in various aspects of their work, including their pay, productivity, safety and comfort, upward mobility, and autonomy on the job. Although there is not a clear hypothesis about which of these factors of employer investment will prove the strongest link with worker attitudes toward automation, survey results can generate new hypothesis about the types of employer investments that are most related to workers’ automation attitudes.
    
    Work environments also can vary sharply by country based on national regulation, training systems, and employment norms. The survey’s multinational sample including nine countries with varied regulatory patterns invites several hypotheses probing cross-national comparison of automation attitudes. Hypothesis 1F is that workers in countries with a strong social safety net –- what political economy scholarship has referred to as coordinated market economies –- will be more positive about the ability of technologies to improve job security. By contrast, workers in countries with strong market principles –- so-called liberal market economies –- will be more positive about the ability of new technologies to influence mobility and ergonomics (Hypothesis 1G). Given the association between liberal market economies like the United States and the United Kingdom, the implication of Hypotheses 1F and 1G is that these countries will report more benefits of automation for work overall. The test for these hypotheses will be if a binary variable indicating workers in liberal market economies or coordinated market economies is associated with workers' perception of the impact of automation (in general) and specifically on job security.

\subsection{Technology interactions}
    
    A worker’s exposure to advanced technologies in their own jobs could reasonably affect how they think about the impact of new technologies on work. For example, if workers have previous experience with automation technologies at work, they may assess the impact of automation based on whether that previous technology succeeded or failed. 
    
    While the survey does not directly measure whether a worker's past experiences with technology have been positive or negative, a rougher measure of general technology exposure is captured, and its impact on workers' perceived benefits of automation is explored (H2A). It is not immediately clear how workers with more exposure to advanced technologies will think about the benefits of automation at work, therefore any findings linking technology exposure and attitudes towards automation are hypothesis generating.
    
    How workers interact with technologies as part of their job may shape their attitudes toward automation. In particular, the survey asks workers about how they typically use technology. Do they troubleshoot, design technology for others, or use the tools they are given? Hypothesis 2B is that workers who troubleshoot technologies as part of their jobs will be more likely to see benefits from automation. The survey also asks about what information workers seek to obtain about the technologies they use. Hypothesis 2C is that workers who say that they need to learn how technology works in detail (understand ``what's under the hood'') are more likely to see benefits from automation, whereas workers who approach technology as a black box and only need it to complete a task are more likely to see the impact of technology on their job as negative. The tests for these hypotheses is whether workers who troubleshoot and workers who need to understand technologies in detail as part of their jobs are also more likely to see benefits from automation for their work, controlling for other factors.
    
    The final hypothesis focuses on the questions about the role that workers play when their employer implements new technologies. Is the worker teaching others how to use the technology? Are they championing it as part of their work? The expectation in Hypothesis 2D is that workers that identify as internal advocates or champions of new technologies will also be optimistic about the impact of new technology on various aspects of their work. The purpose of this hypothesis is to establish a strong link between willingness to champion a technology and the perceived impact of automation on the quality of work. If this link is confirmed, we can infer that workers who say they advocate for technology and workers who are optimistic about technology are largely the same group.

\subsection{Attitudes toward work and automation}

    Previous studies of automated systems have shown that an individual’s level of trust –- as measured through a series of questions –- can predict their openness to new technologies. Hypothesis 3A is that a worker’s trust –- as measured through the same questions –- is positively associated with the worker’s attitudes toward automation.
    
    The survey also measures worker job satisfaction across multiple dimensions. Hypothesis 3B is that a worker’s job satisfaction is associated with a worker’s perception of the impact of automation on their jobs. The direction of the linkage between job satisfaction and perceptions of automation is not immediately clear, however. One perspective is that workers who are more satisfied with their jobs will be more optimistic about potential changes to it due to new technologies. A second perspective is that those satisfied with their jobs will be resistant to change. And by contrast, those with lower satisfaction will be more likely to be optimistic about the possibilities of new technology. The test for this hypothesis will be whether a composite measure of job satisfaction (the average of workers' responses to several questions about their job satisfaction) is associated with the composite measure of how workers perceive automation affects various aspects of their work.
    
    Similarly, hypothesis 3C predicts a link between what workers value in their jobs –- what factors they report contribute to a good job –- and their perceptions of automation. It is not obvious how workers’ conception of a good job will align with their perceptions of automation. It could be that the two factors are complementary: workers who value autonomy at work, for example, may also perceive automation to provide benefits for autonomy at work. However, this linkage appears to depend on workers’ underlying attitudes toward automation. Thus, it is possible that what workers perceive as important to a good job is a mediating factor for how workers perceive the impact of automation on their work. 

\subsection{Organizational incentives}

    The embedded experiment in the survey is designed to test two separate hypotheses. The experiment has two treatment groups, one corresponding to each hypothesis. The first treatment group is told that they will –- as workers –- be able to provide input on how a hypothetical technology is implemented within their firm. Hypothesis 4A, arising from research on work and organization, is that workers will perceive automation more positively if they have voice in how it is implemented. The inclusion of an opportunity for worker input is the sole difference between the control group and the first treatment group. 
    
    The second treatment group is told that they will receive financial incentives –- a bonus –- if they improve their productivity after the introduction of a hypothetical technology. In field research, we have seen companies adopt incentives related to their productivity. Hypothesis 4B suggests that financial incentives can induce a more positive outlook toward technological change, turning potential resisters of a technology into champions. In the survey experiment, the sole difference between the control and the second treatment group is the mention of financial incentives. 

    The ``treatment effect'' in both cases is the difference between how workers faced with each treatment condition perceive the potential impact of the technology Alpha on various aspects of their work.

\subsection{Technology appearance}

    Previous work in human-robot interaction has underscored an importance for the appearance of technology to match the severity of a task~\cite{manja2008-domestic, goetz2003-matching, zlotowski2020-one}. We test two hypotheses to better understand how technology design affects how workers perceive the impact of automation and new technology in work environments. 

    Hypothesis 5A is that someone's work environment is associated with their perceptions of individual technologies. In particular, the prediction is that workers in production and warehouse settings --- more severe environments --- and who might be more familiar with machine-like appearances are more likely to report that the more machine-like technology designs are useful. We test this hypothesis in a binomial logit model with a binary outcome variable (is the technology considered useful?), and a binary independent variable (is the worker in a production or warehousing work environment?). 
    
    Hypothesis 5B is that workers overall will be more likely to identify machine-like technologies as useful in more severe environments (e.g. construction), and less machine-like technologies as more useful in less severe environments (e.g. restaurants, hospitals). The statistical test for this hypothesis will be the difference in perceived usefulness of two technology designs given different work environments. 

\section{What Workers Think About Automation: Descriptive Statistics}

Before asking why some workers are more open to automation and technological change than others, there are more basic questions: how many workers are open to technological change in the first place? What share of workers feel valued on the job? What share report performing routine tasks at work? How many are early adopters of new technologies? On some of these questions, there is an opportunity to draw a statistical baseline to mark how workers think about automation. Since survey response patterns can vary by country -- and hypotheses are primarily tested at the country level -- this section presents the data by country across more than twenty questions in the survey. It focuses on the questions that are key to the hypotheses stated above, and a broader set of statistical tables are available in the Data Appendix.

Table 1 describes the work environment of the sample of workers that perform jobs in occupational categories that we classify as office jobs (e.g. marketing and sales, finance, etc.). It also identifies the share of workers by country that say they frequently engage in routine tasks or frequently engage in complex problem-solving. On a five-point Likert scale, the respondents that frequently engaged in these tasks selected 4 (Agree) or 5 (Strongly Agree). The table also reports the percentage of workers who said they were satisfied with different aspects of their job, as well as workers who signaled that they were among the earliest to use new technologies.

\begin{table}[h]
\centering
\resizebox{\columnwidth}{!}{
\begin{tabular}{llllllll}
\toprule
Country & Workers & Avg Age & Office jobs & Routine & Problem-solving & Job satisfaction & Early tech adopter \\ \hline \\
Australia      & 502  & 42 & 81\%& 56\%& 30\%& 51\%& 29\%\\
France         & 506  & 43 & 84\%& 41\%& 26\%& 46\%& 34\%\\
Germany        & 505  & 44 & 84\%& 44\%& 26\%& 61\%& 27\%\\
Italy          & 511  & 44 & 81\%& 44\%& 37\%& 42\%& 44\%\\
Japan          & 510  & 45 & 84\%& 52\%& 17\%& 19\%& 23\%\\
Poland         & 500  & 40 & 78\%& 47\%& 25\%& 50\%& 37\%\\
Spain          & 500  & 43 & 80\%& 46\%& 28\%& 52\%& 42\%\\
United Kingdom & 501  & 42 & 83\%& 54\%& 30\%& 56\%& 32\%\\
United States  & 4833 & 47 & 80\%& 49\%& 27\%& 64\%& 30\%\\ \hline
\end{tabular}
}
\caption{Worker Characteristics by Country}
\label{tab:my-table}
\end{table}

Table 2 identifies how workers evaluate different factors as contributors to the enjoyment of their work --- what makes a good job. It also shows how workers see their employer as invested in various aspects of their jobs. In these data, the percentage notes the share of workers who noted this factor as very important (4 or 5 on a five-point Likert scale). In comparing the baseline by country, it is useful to compare data \textit{within} countries, not between countries. For example, a smaller share of Japanese workers identify pay as an important factor to enjoying their work than any other country. However, for Japanese workers, this factor is still 10 percentage points more important than any other factor listed.

\begin{table}[h]
\centering
\resizebox{\columnwidth}{!}{
\begin{tabular}{llllllllll}
\toprule
        &         & \multicolumn{4}{l}{Factors important to enjoying your work} & \multicolumn{4}{l}{Factors your employer is invested in} \\ \hline \\
Country & Workers & Safety \& comfort   & Pay  & Autonomy  & Upward mobility  & Safety \& comfort  & Pay  & Autonomy  & Upward mobility  \\ \hline \\
Australia      & 502  & 62\%& 69\%& 54\%& 45\%& 36\%& 16\%& 35\%& 25\%\\
France         & 506  & 55\%& 59\%& 46\%& 40\%& 29\%& 23\%& 27\%& 29\%\\
Germany        & 505  & 54\%& 62\%& 57\%& 44\%& 31\%& 25\%& 45\%& 29\%\\
Italy          & 511  & 56\%& 58\%& 45\%& 47\%& 40\%& 25\%& 41\%& 37\%\\
Japan          & 510  & 42\%& 52\%& 29\%& 26\%& 17\%& 15\%& 15\%& 14\%\\
Poland         & 500  & 56\%& 68\%& 43\%& 47\%& 35\%& 27\%& 29\%& 28\%\\
Spain          & 500  & 60\%& 64\%& 50\%& 50\%& 32\%& 23\%& 35\%& 27\%\\
United Kingdom & 501  & 62\%& 64\%& 56\%& 48\%& 37\%& 24\%& 40\%& 30\%\\
United States  & 4833 & 60\%& 70\%& 57\%& 40\%& 34\%& 20\%& 45\%& 29\
\\ \hline
\end{tabular}
}
\caption{Job Attitudes by Country}
\label{tab:my-table}
\end{table}

Finally, Table 3 reports descriptive statistics of the outcomes of interest: the impact of automation on various dimensions of work, including safety and comfort, pay, autonomy (the ability for workers to determine what they do and how they do it), and upward mobility (learning and growing in their careers). The percentages are reported as the net impact. Given that the five-point Likert scales range from negative to positive with a score of 3 as neutral, we calculate the net impact as the difference between the positive share (workers reporting four or five on the Likert scale) and the negative share (workers reporting one or two on the Likert scale). A negative net impact indicates more workers report a negative impact than report a positive impact.

Four findings from these descriptive data stand out.

\begin{enumerate}
    \item First, the overall share of workers who identify as early adopters of new technologies is low, hovering around a third of workers across the nine countries in the sample. Moreover, the share of workers that say they frequently engage in problem-solving as part of their jobs is similarly low (see Table 1).
    
    \item Second, although the United States has long been associated with opportunities for intergenerational mobility, American workers report that opportunities to learn and grow in their careers --- which we describe as "upward mobility" --- is the least significant factor when workers think about what makes a good job (see Table 2). American workers are among the least likely to report that gaining new skills and growing in their careers is highly important to the quality of their job.

    \item Third, the perceived impact of automation on various aspects of work is broadly positive with the sole negative perceptions being in the United States for the impact of automation on pay and job security (see Table 3). By contrast, a strong plurality of people perceive the impact of automation and new technology to be positive for safety and comfort, autonomy, and upward mobility at work.

    \item Fourth, despite American workers having the highest job satisfaction among all countries in the survey, American workers are most pessimistic about the impact of automation on three of the five dimensions. Given the reputation of the United States as a center of technological innovation and a supporter of innovative advances, this pessimism toward the impact of technology on work is surprising.
\end{enumerate}

\begin{table}[H]
\centering
\resizebox{\columnwidth}{!}{
\begin{tabular}{lllllll}
\toprule
               &         & \multicolumn{5}{l}{Net impact of new technology \& automation on your work} \\ \hline \\ 
Country        & Workers & Safety \& comfort   & Pay    & Autonomy  & Upward mobility  & Job security  \\ \hline \\
Australia      & 502     & 32\%               & 4\%   & 23\%     & 37\%            & 1\%          \\
France         & 506     & 29\%               & 5\%   & 16\%     & 23\%            & 9\%          \\
Germany        & 505     & 33\%               & 10\%  & 20\%     & 35\%            & 11\%         \\
Italy          & 511     & 46\%               & 18\%  & 40\%     & 36\%            & 14\%         \\
Japan          & 510     & 34\%               & 7\%   & 16\%     & 25\%            & 2\%          \\
Poland         & 500     & 48\%               & 17\%  & 25\%     & 43\%            & 49\%         \\
Spain          & 500     & 37\%               & 13\%  & 30\%     & 37\%            & 20\%         \\
United Kingdom & 501     & 29\%               & 3\%   & 19\%     & 34\%            & 1\%          \\
United States  & 4833    & 30\%               & -1\%  & 14\%     & 35\%            & -5\%  \\ \hline
\end{tabular}
}
\caption{\\ Perceived impact of automation by country}
\label{tab:my-table}
\end{table}

\section{Results}

For a survey this broad in scope, it should come as no surprise that there is not one dominant factor that helps explain why some workers are more open to automation than others. The majority of the hypothesized factors have some association with workers' outlook on automation --- and apparent willingness to champion technological change. This section reviews the evidence for each hypothesis in two ways. It presents summary data for the hypothesized variables in crosstabular format, in addition to non-parametric tests where the hypothesized factors in question are independent variables and individuals' perceptions of the impact of automation is the dependent variable. Where possible the analysis aims to test the hypotheses using multiple methods to discern whether the statistical relationships in question are robust to multiple specifications. 

For each category of hypotheses, the results section summarizes the hypotheses, the variables used to operationalize them, and the result in a summary table. The summary tables also reference the part of the table or figure on which the result is based. In many cases, the results serve as a test of the hypotheses presented, as well as a generator of new hypotheses. This section begins to explore potential explanations of the findings from the survey.

\subsection{Work environment}

Although new technologies are often presented as “skill-biased,” benefiting more educated workers, perceptions of automation in some cases follow the opposite pattern. Workers with less formal education are often more optimistic about the impact of automation on their job security. In Table 6, there are several conditions under which worker's education -- measured by a variable with four categories including less than high school, high school, some college, and bachelors degree or higher -- is negatively and significantly associated with their perception of the impact of automation.\footnote{There are two important features of the education variable in the survey. First, it is different for each country, so the analysis of the education variable in this case focuses exclusively on the United States. Second, the association of education with the perceived impact of automation is sensitive to how the variable is defined. If it is a binary variable where college is 1 and non-college is 0, the association is weak or not statistically significant. If it is treated as a continuous variable with four values with roughly equal intervals (the categories are proxies for years of education, which is a continuous variable), then the association between education and the perceived impact of automation is negative and significant.}

\begin{table}[]
\begin{tabular}{|m{0.05\linewidth} | m{0.33\linewidth} | m{0.26\linewidth}  | m{0.25\linewidth} |}
 \hline
 Hyp & \makecell{Description} & \makecell{Measure} & \makecell{Result} \\ 
 \hline
H1A &
  \makecell{Workers who are more educated\\ will be more favorable\\ toward the impact of automation\\ technologies at work.} &
  \makecell{Binary variable of education\\ (college / non-college)} &
  \makecell{Hypothesis rejected. \\ 
  \textbf{Table 6, cols 1-3} \\ 
  Row: Education}\\
  \hline
H1B &
  \makecell{Workers who perform more routine \\ tasks will view automation \\ technologies less favorably.} &
  \makecell{Binary variable \\ of routine tasks\\ (frequent, not frequent)} &
  \makecell{ Mixed results. \\ Frequency of problem \\solving tasks \\ more significant. \\ 
  \textbf{Table 6, cols 1-5} \\
  Rows: Routine tasks \\ 
  Problem solving } \\
  \hline
H1C &
  \makecell{Workers will be more favorable \\ toward technologies that improve their \\ safety and comfort,\\ or their autonomy at work.} &
  \makecell{Descriptive statistics of \\ “automation impact” score\\  on safety and comfort,\\ as well as autonomy, at work.} & 
  \makecell{Hypothesis supported. \\ 
  \textbf{Table 3, cols 3,5}} \\
  \hline
H1D &
  \makecell{Workers will be less favorable \\ toward technologies\\ that threaten their job security.} &
  \makecell{Descriptive statistics of \\ perceived “automation \\ impact” score on job security.} &
  \makecell{Evidence mixed. \\ Perceived impact \\ of automation on job\\ security has lowest net score. \\ Workers in every country \\ except the U.S. perceives \\ net positive impact\\ of automation on job security. \\ 
  \textbf{Table 3, col 7}} \\
  \hline
H1E &
  \makecell{Workers who see their employer\\ as invested in them will be\\ more favorable toward automation.} &
  \makecell{Binary variable of employee \\ feeling valued by employer.} &
  \makecell{Hypothesis supported. \\
  \textbf{Table 6, cols 2-5} \\
  Row: valued
  by employer} \\
  \hline
H1F &
  \makecell{Workers in countries with a strong \\ safety net (coordinated market \\ economies) will be more \\ positive about the \\ ability of technologies to \\ improve job security.} &
  \makecell{Binary variable of strong \\ social safety net countries\\ (France, Germany,\\ Italy, Spain)r} &
  \makecell{Mixed results. \\ Descriptive data \\ indicate hypothesis. \\ Regressions do not \\ support hypothesis. \\ 
  \textbf{Table 3, cols 3, 6} \\
  Rows: France, Germany \\ 
  Italy, Spain \\ 
  See also \textbf{Table 6, col 4} \\
  Row: Coord. Market Econ. \\ 
} \\
\hline
H1G &
  \makecell{Workers in countries with strong \\ market principles will be \\ more positive about the\\ ability of new technologies\\ to influence mobility and ergonomics} &
  \makecell{Binary variable of \\ Liberal Market Economies\\ (Australia, U.K., U.S.)} &
  \makecell{Hypothesis supported.\\
  \textbf{Table 6, Col 5} \\ 
  Row: Lib. Market Econ.} \\* \midrule

\end{tabular}
\caption{Summary of work environment hypotheses and results}
\label{table:1}
\end{table}

To test hypotheses related to the tasks workers perform on the job, we included the variables "routine" and "complex problem-solving" as binary measures of the frequency with which workers perform each task. The findings in the regression models are consistent with the descriptive statistics. They indicate workers performing jobs that require complex problem solving or new ideas tend to be more positive about automation. 

Although economists have long identified workers in routine jobs as most vulnerable to automation, workers’ level of routine tasks does not seem to be the strongest predictor of how they think about the impact of automation on their work. Instead, workers report doing a variety of tasks as part of their jobs with complex problem-solving tasks as among the strongest predictors of workers seeing benefits from new technologies.

\begin{table}[]
\centering
\resizebox{\columnwidth}{!}{
\begin{tabular}{llllllll}
\toprule
         \multicolumn{8}{c}{Net impact of automation \& new technology on your work} \\ \hline \\
Race     & Education  & Workers & Safety \& comfort  & Pay   & Autonomy  & Upward mobility  & Job security  \\
\hline \\ 
Black    & no college & 263     & 35\%                & 9\%  & 17\%     & 33\%            & 14\%         \\
Hispanic & no college & 373     & 31\%                & 16\% & 26\%     & 32\%            & 15\%         \\
Other    & no college & 113     & 27\%                & 3\%  & 5\%      & 19\%            & -5\%         \\
White    & no college & 1531    & 27\%                & -6\% & 6\%      & 24\%            & -8\%         \\
Black    & college    & 184     & 46\%                & 9\%  & 20\%     & 46\%            & 3\%          \\
Hispanic & college    & 201     & 31\%                & 5\%  & 21\%     & 39\%            & 1\%          \\
Other    & college    & 257     & 35\%                & -1\% & 15\%     & 44\%            & -9\%         \\
White    & college    & 1935    & 30\%                & -3\% & 16\%     & 43\%            & -9\%\\ \hline   
\end{tabular}
}
\caption{Perceived impact of automation by race and education}
\label{tab:my-table}
\end{table}

\begin{table}[]
\centering
\resizebox{\columnwidth}{!}{
\begin{tabular}{lllllllll}
\toprule
&            &     &    & \multicolumn{5}{l}{Net impact of automation \& new technology on your work} \\ \hline \\
Job type &
  \begin{tabular}[c]{@{}l@{}}Routine \\    tasks\end{tabular} &
  \begin{tabular}[c]{@{}l@{}}Problem\\    solving\end{tabular} &
  Workers &
  Safety &
  Pay &
  Autonomy &
  Upward mobility &
  Job security \\ \hline \\
Office   & low  & low  & 2066 & 31\%& 5\% & 21\%& 37\%& 5\% \\
Physical & low  & low  & 562  & 33\%& 1\% & 18\%& 33\%& 2\% \\
Office   & high & low  & 1692 & 30\%& -6\%& 11\%& 28\%& -5\%\\
Physical & high & low  & 736  & 32\%& -7\%& 7\% & 17\%& -6\%\\
Office   & low  & high & 855  & 39\%& 14\%& 28\%& 48\%& 9\% \\
Physical & low  & high & 168  & 41\%& 16\%& 24\%& 47\%& 12\%\\
Office   & high & high & 678  & 43\%& 25\%& 33\%& 45\%& 25\%\\
Physical & high & high & 242  & 33\%& 27\%& 25\%& 33\%& 21.5\%\\ 
\hline 
\end{tabular}
}
\caption{Perceived net impact of automation by job type}
\label{tab:my-table}
\end{table}

In examining other factors that may influence perceptions of automation, we also included demographic variables such as gender, age, and race in our models. We identified that Black and Hispanic workers in the United States are far more optimistic about the impact of automation on their jobs than workers of other racial and ethnic backgrounds. This was an unexpected finding without a clear explanation. The apparent optimism of African American and Hispanic workers about automation invites new hypotheses that can be tested in future research.

Workers who feel like their employer values them and is invested in their safety are more optimistic about the impact of automation at work. Workers’ job satisfaction and level of trust are also highly predictive of their attitudes toward automation.

\begin{table}[!htbp] \centering 
  \caption{Predictors of Worker Attitudes toward Automation} 
  \label{} 
\footnotesize 
\begin{tabular}{@{\extracolsep{1pt}}lccccc} 
\\[-1.8ex]\hline 
\hline \\[-1.8ex] 
 & \multicolumn{5}{c}{\textit{Dependent variable:}} \\ 
\cline{2-6} 
\\[-1.8ex] & \multicolumn{5}{c}{Positive impact of automation on own job} \\ 
\\[-1.8ex] & (1) & (2) & (3) & (4) & (5)\\ 
\hline \\[-1.8ex] 
 Age & $-$0.001 (0.001) & $-$0.002$^{**}$ (0.001) & $-$0.001 (0.001) & 0.002$^{***}$ (0.001) & 0.002$^{***}$ (0.001) \\ 
  Sex (Female) & 0.02 (0.02) & 0.04$^{**}$ (0.02) & 0.04$^{*}$ (0.02) & 0.03$^{*}$ (0.02) & 0.03$^{*}$ (0.02) \\ 
  Full-time & 0.06$^{**}$ (0.03) & 0.05$^{*}$ (0.03) & 0.04 (0.03) & 0.01 (0.02) & 0.01 (0.02) \\ 
  Trust & 0.11$^{***}$ (0.01) & 0.05$^{***}$ (0.01) & 0.04$^{***}$ (0.01) & 0.05$^{***}$ (0.01) & 0.05$^{***}$ (0.01) \\ 
  Education & $-$0.01 (0.01) & $-$0.05$^{***}$ (0.01) & $-$0.04$^{***}$ (0.01) &  &  \\ 
  Income & 0.01 (0.01) & $-$0.004 (0.01) & $-$0.001 (0.01) &  &  \\ 
  Physical jobs & $-$0.03 (0.03) & $-$0.02 (0.03) & $-$0.02 (0.03) & $-$0.01 (0.02) & $-$0.01 (0.02) \\ 
  Routine tasks & $-$0.05$^{**}$ (0.02) & $-$0.03 (0.02) & $-$0.04$^{*}$ (0.02) & 0.003 (0.02) & 0.005 (0.02) \\ 
  Problem solving & 0.21$^{***}$ (0.02) & 0.08$^{***}$ (0.02) & 0.05$^{*}$ (0.02) & 0.05$^{***}$ (0.02) & 0.05$^{***}$ (0.02) \\ 
  Job satisfaction &  & 0.16$^{***}$ (0.01) & 0.12$^{***}$ (0.02) & 0.10$^{***}$ (0.01) & 0.10$^{***}$ (0.01) \\ 
  Valued by employer &  & 0.13$^{***}$ (0.02) & 0.10$^{***}$ (0.03) & 0.09$^{***}$ (0.02) & 0.09$^{***}$ (0.02) \\ 
  Tech adoption &  & 0.14$^{***}$ (0.02) & 0.12$^{***}$ (0.02) & 0.09$^{***}$ (0.01) & 0.09$^{***}$ (0.01) \\ 
  Video games &  & $-$0.02$^{***}$ (0.01) & $-$0.02$^{***}$ (0.01) & $-$0.01$^{*}$ (0.01) & $-$0.01$^{*}$ (0.01) \\ 
  Screen time &  & 0.04$^{***}$ (0.01) & 0.04$^{***}$ (0.01) & 0.04$^{***}$ (0.01) & 0.04$^{***}$ (0.01) \\ 
  Early adopter &  & 0.11$^{***}$ (0.02) & 0.11$^{***}$ (0.02) & 0.11$^{***}$ (0.02) & 0.11$^{***}$ (0.02) \\ 
  Champion &  & 0.23$^{***}$ (0.02) & 0.21$^{***}$ (0.02) & 0.21$^{***}$ (0.02) & 0.21$^{***}$ (0.02) \\ 
  Troubleshooter &  & $-$0.07$^{**}$ (0.03) & $-$0.08$^{***}$ (0.03) & $-$0.03 (0.02) & $-$0.03 (0.02) \\ 
  Learner &  & 0.10$^{***}$ (0.02) & 0.08$^{***}$ (0.02) & 0.06$^{***}$ (0.02) & 0.06$^{***}$ (0.02) \\ 
  Empl inv prod. &  &  & 0.08$^{***}$ (0.03) & 0.08$^{***}$ (0.02) & 0.08$^{***}$ (0.02) \\ 
  Empl inv pay &  &  & $-$0.002 (0.03) & 0.02 (0.02) & 0.02 (0.02) \\ 
  Emp inv ergonomics &  &  & 0.08$^{***}$ (0.03) & 0.08$^{***}$ (0.02) & 0.08$^{***}$ (0.02) \\ 
  Emp inv autonomy &  &  & $-$0.01 (0.02) & $-$0.02 (0.02) & $-$0.02 (0.02) \\ 
  Emp inv mobility &  &  & 0.05 (0.03) & 0.06$^{***}$ (0.02) & 0.06$^{***}$ (0.02) \\ 
  Imp. of pay &  &  & 0.01 (0.03) & $-$0.003 (0.02) & $-$0.003 (0.02) \\ 
  Imp. of ergonomics &  &  & 0.01 (0.02) & 0.03$^{*}$ (0.02) & 0.03$^{*}$ (0.02) \\ 
  Imp. of autonomy &  &  & 0.03 (0.02) & 0.01 (0.02) & 0.01 (0.02) \\ 
  Imp. of mobility &  &  & 0.08$^{***}$ (0.02) & 0.05$^{***}$ (0.02) & 0.05$^{***}$ (0.02) \\ 
  Coord. Market Econ. &  &  &  & 0.003 (0.02) &  \\ 
  Lib. Market Econ. &  &  &  &  & $-$0.06$^{**}$ (0.02) \\ 
  Constant & 2.84$^{***}$ (0.07) & 2.24$^{***}$ (0.09) & 2.30$^{***}$ (0.09) & 2.15$^{***}$ (0.06) & 2.16$^{***}$ (0.06) \\ 
 \hline \\[-1.8ex] 
Observations & 4,951 & 4,880 & 4,834 & 8,869 & 8,869 \\ 
R$^{2}$ & 0.04 & 0.15 & 0.16 & 0.17 & 0.17 \\ 
Adjusted R$^{2}$ & 0.04 & 0.15 & 0.15 & 0.17 & 0.17 \\ 
Residual Std. Error & 0.76 (df = 4941) & 0.71 (df = 4861) & 0.70 (df = 4806) & 0.70 (df = 8842) & 0.70 (df = 8842) \\ 
\hline 
\hline \\[-1.8ex] 
& \multicolumn{5}{r}{$^{*}$p$<$0.1; $^{**}$p$<$0.05; $^{***}$p$<$0.01} \\ 
& \multicolumn{5}{r}{Note: Columns 1-3 are U.S.-specific models. Columns 4-5 are global models.}
\end{tabular} 
\end{table} 

\subsection{Technology interactions}

\begin{table}[!htb]
\begin{tabular}{|m{0.05\linewidth} | m{0.33\linewidth} | m{0.26\linewidth}  | m{0.25\linewidth} |}
 \hline
 Hyp & \makecell{Description} & \makecell{Measure} & \makecell{Result} \\ 
 \hline
H2A &
  \makecell{Workers with more experience \\ working in a technologically \\ advanced environment will \\ have a different \\ perception of automation.} &
  \makecell{Index of employer adoption \\ of robotics, AI, and \\generative AI (mean of\\ frequencies that workers \\reported their companies using\\ these technologies).} &
  \makecell{Exposure to technology \\ at work associated \\ with positive perceptions \\ of the impact of \\ automation at work. \\ 
  \textbf{Table 6, cols 2-5} \\ 
  Row: Tech adoption } \\
  \hline
H2B &
  \makecell{Workers who troubleshoot \\ technologies as part of\\ their jobs will be more likely\\  to see benefits from automation} &
  \makecell{Binary variable of \\whether workers \\ troubleshoot technology.} & \makecell{
  Hypothesis not supported. \\ 
  Evidence of reverse. \\ 
  \textbf{Table 6, cols 2-3} \\ 
  Row: Troubleshooter \\ } \\
  \hline
H2C &
  \makecell{Workers who are motivated\\ to learn more about \\ technologies are more likely\\ to see benefits from automation.} &
  \makecell{Binary measure of worker’s\\ need to learn about\\ technologies at work.} &
  \makecell{Hypothesis supported. \\ 
  \textbf{Table 6, cols 2-5} \\ 
  Row: Learner} \\
  \hline
H2D &
  \makecell{Workers that identify as advocates\\ of new technologies will \\ be optimistic about the impact of\\ technology on their work.} &
  \makecell{Binary measure of whether\\ workers identify as a strong\\ advocate for the use \\ of technology at work} &
  \makecell{ Hypothesis supported. \\ 
  \textbf{Table 6, cols 2-5} \\ 
  Row: Champion} \\
  \hline
\end{tabular}
\caption{Summary of technology interactions hypotheses and results}
\label{table:technology}
\end{table}

We anticipated that a worker's previous experiences with advanced technologies at work, as well as the way they interact with technology in their particular role, will shape how they perceive the impact of technologies on their job. The results indicate that workers with more exposure to new technologies like AI and robotics in their jobs indeed report that the impact of new technologies and automation on their jobs will be more positive. This finding was not a foregone conclusion. 

It is possible that workers with more exposure to these technologies might be more pessimistic about the impact of new technologies on their work, but this was not the case. It is not immediately clear why more experience with new technologies is linked to optimism. It could be that workers' prior experiences with new technologies have been positive and made them optimistic. An alternative is that workers' previous experiences have convinced them that new technologies are not threatening to their job, pay, or other features of their work that they like. A third option is that workers who are more optimistic about the impact of new technologies have found roles where they are more exposed to those technologies at work.

Several dimensions of workers' interactions with technologies at work are associated with their perceptions of the impact of new technologies. First, if a worker self-identifies as an advocate or champion of a new technology on the job, it is a strong predictor of whether they see new technologies and automation as beneficial for their job. This strong link is consistent with our hypothesis that champions of new technologies see benefits of new technologies for their work. 

Other dimensions of how workers interact with technology on the job were less clearly related. We anticipated that workers who troubleshoot technology as part of their work might be more optimistic about the impact that technology will have on their jobs, perhaps because the troubleshooting workers realize that human intelligence is required to make technology work. However, these workers might also experience the frustrations of technology more readily and see these as a negative impact on their job quality. The result in the survey was that workers who troubleshoot are no more optimistic about new technologies than other workers after controlling for other variables.

Workers who must ``go under the hood'' of a new technology to understand how it works are significantly more likely to see technology as beneficial for their jobs. Why might the requirement to learn about new technologies be more predictive of one's attitude toward automation than the need to troubleshoot new technologies? One potential explanation is that troubleshooting can be an involuntary and negative interaction with technology, whereas curiosity about what is going on under the hood of a technology, which is positively associated with openness to automation, is framed as part of learning.

\subsection{Attitudes toward work and automation}

The survey asked workers a series of questions about their attitudes toward work and technology, as well as their outlook in general. For example, a series of questions about workers' willingness to trust was included in the survey because it has been previously found to predict how individuals interact with new technologies. 

In this survey, an individual's willingness to trust, as measured by an average of their responses to the battery of trust questions, is positively and significantly associated with how they perceive the impact of automation on their work (Hypothesis 3A). In other words, workers who are more trusting are also more likely to see benefits of automation for their jobs. 

Worker's self-reported job satisfaction followed a similar pattern. High job satisfaction was positively and significantly associated with a worker's perception of the impact of automation on their work (hypothesis 3B). This finding is interesting because it could be that workers who are satisfied in their jobs report negative feelings about the potential impact of technology to disrupt their work. But this survey finds the opposite. Workers who are more satisfied in their jobs -- as well as workers who feel more valued by their employers --- are also more likely to see potential benefits of automation and new technologies for their work.

There was no clear prediction about how workers' enjoyment of their work --- the aspects of their work that they found more or less important for their job quality --- would be related to their perceived impact of automation. Hypothesis H3C suggested that workers that valued some aspects of their work may have a more positive perception of automation than workers that valued other aspects of their work. The results found that workers that valued learning and upward mobility were also more likely to see potential benefits from automation and new technologies, even after controlling for covariates. This finding makes sense in context because workers also reported a strongly positive perceived impact of automation on learning and upward mobility.

\begin{table}[!htb]
\begin{tabular}{|m{0.05\linewidth} | m{0.33\linewidth} | m{0.26\linewidth}  | m{0.25\linewidth} |}
 \hline
 Hyp & \makecell{Description} & \makecell{Measure} & \makecell{Result} \\ 
 \hline
H3A &
  \makecell{Worker’s propensity to trust is\\ associated with worker’s\\ attitudes toward automation.} &
  \makecell{Index variable including \\ workers’ answers to a \\ series of questions about \\ their willingness to trust.} &
  \makecell{Hypothesis supported. \\ 
  \textbf{Table 6, cols 1-5} \\ 
  Row: Trust } \\
  \hline
H3B &
  \makecell{Worker’s job satisfaction is\\ associated with a worker’s\\ perception of the impact of\\ automation on their jobs} &
  \makecell{Index of how satisfied  \\workers are with various  \\aspects of their jobs.} &
  \makecell{Hypothesis supported. \\ 
  \textbf{Table 6, cols 2-5} \\ 
  Row: Job satisfaction} \\
  \hline
H3C &
  \makecell{What workers value as part\\ of their enjoyment of their\\ job is associated with their\\ perception of automation.} &
  \makecell{Binary variables of what\\ workers value (safety and\\ comfort, pay, autonomy,  \\upward mobility).} &
  \makecell{ Hypothesis supported. \\
  Positive association \\ 
  with valuing upward mobility. \\
  \textbf{Table 6, cols 3-5} \\ 
  Row: Imp. of mobility } \\
  \hline
\end{tabular}
\caption{Summary of work attitudes hypotheses and results}
\label{table:work_attitudes}
\end{table}

\subsection{Organizational incentives}

Extensive research on worker engagement has suggested that new opportunities for worker voice can improve the effectiveness of technology adoption. However, in our experiment, we did not find significant impact of worker voice on how workers perceived the new technology Alpha. Workers who were offered an opportunity to shape the new technology saw it much the same way as workers who did not.

The null result of the "worker voice" treatment has multiple potential explanations. The first is that the treatment was too weak. Participants in the experiment merely read about a potential technology and were told that they would have an opportunity to provide input. They could have read this opportunity to provide input as insignificant. By contrast, if they were actually given the chance to provide input, they may perceive automation as more beneficial as a result. One supporting piece of evidence is that workers that feel more valued by their employer --- which may be a proxy for feeling as if you provide valued input --- are also more likely to perceive benefits from automation for their work.

The second potential explanation is that only a small subsection of workers value the ability to provide input about their work environment. Other workers are more focused on other aspects of their jobs, such as their safety and comfort at work or their pay. The supporting evidence for this explanation is the relatively high share of workers who reported pay and safety and comfort as important for the enjoyment of their work --- a higher share than reported their ability to control what they do and how they do it (their autonomy) at work.

The second treatment --- a potential financial bonus for workers whose productivity increased after using the technology --- elicited a different response. Workers offered a financial incentive had a significantly more positive perception of how the technology Alpha would affect their job. The treatment effect was positive not only for the prospective impact of Alpha on workers' pay. It was also positive and statistically significant for the prospective impact of Alpha on workers' job security. Why might workers think that financial incentives for using automation could translate into more benefits for their job security?

\begin{table}[!htbp] \centering 
  \caption{Experimental Results} 
  \label{} 
\footnotesize 
\begin{tabular}{@{\extracolsep{1pt}}lcccc} 
\\[-1.8ex]\hline 
\hline \\[-1.8ex] 
 & \multicolumn{4}{c}{\textit{Dependent variable:}} \\ 
\cline{2-5} 
\\[-1.8ex] & \multicolumn{4}{c}{Positive impact of automation on own job} \\ 
\\[-1.8ex] & (1) & (2) & (3) & (4)\\ 
\hline \\[-1.8ex] 
 Treatment (Voice) & $-$0.03 (0.02) &  & $-$0.03$^{*}$ (0.02) &  \\ 
  Treatment (Bonus) &  & 0.05$^{**}$ (0.02) &  & 0.05$^{***}$ (0.02) \\ 
  Age &  &  & 0.001$^{*}$ (0.001) & 0.001 (0.001) \\ 
  Sex (Female) &  &  & 0.02 (0.02) & 0.03 (0.02) \\ 
  Full-time &  &  & 0.01 (0.02) & $-$0.01 (0.02) \\ 
  Trust &  &  & 0.06$^{***}$ (0.01) & 0.05$^{***}$ (0.01) \\ 
  Physical jobs &  &  & $-$0.01 (0.02) & 0.003 (0.02) \\ 
  Routine tasks &  &  & 0.01 (0.02) & 0.001 (0.02) \\ 
  Problem solving &  &  & $-$0.003 (0.02) & $-$0.02 (0.02) \\ 
  Job satisfaction &  &  & 0.06$^{***}$ (0.01) & 0.05$^{***}$ (0.01) \\ 
  Valued by employer &  &  & 0.04$^{*}$ (0.02) & 0.08$^{***}$ (0.02) \\ 
  Tech adoption &  &  & 0.13$^{***}$ (0.01) & 0.13$^{***}$ (0.01) \\ 
  Video games &  &  & $-$0.01 (0.01) & $-$0.01 (0.01) \\ 
  Screen time &  &  & 0.02$^{**}$ (0.01) & 0.02$^{**}$ (0.01) \\ 
  Early adopter &  &  & 0.11$^{***}$ (0.02) & 0.06$^{***}$ (0.02) \\ 
  Champion &  &  & 0.22$^{***}$ (0.02) & 0.24$^{***}$ (0.02) \\ 
  Troubleshooter &  &  & $-$0.03 (0.02) & 0.001 (0.03) \\ 
  Learner &  &  & 0.07$^{***}$ (0.02) & 0.07$^{***}$ (0.02) \\ 
  Empl inv prod. &  &  & 0.06$^{**}$ (0.03) & 0.06$^{**}$ (0.03) \\ 
  Empl inv pay &  &  & 0.04 (0.03) & 0.07$^{**}$ (0.03) \\ 
  Emp inv ergonomics &  &  & 0.08$^{***}$ (0.02) & 0.04 (0.02) \\ 
  Emp inv autonomy &  &  & $-$0.06$^{***}$ (0.02) & $-$0.03 (0.02) \\ 
  Emp inv mobility &  &  & 0.03 (0.03) & 0.03 (0.03) \\ 
  Importance of pay &  &  & $-$0.02 (0.02) & 0.05$^{**}$ (0.02) \\ 
  Importance of ergonomics &  &  & 0.03 (0.02) & 0.02 (0.02) \\ 
  Importance of autonomy &  &  & $-$0.02 (0.02) & $-$0.03 (0.02) \\ 
  Importance of mobility &  &  & 0.05$^{**}$ (0.02) & 0.05$^{**}$ (0.02) \\ 
  Constant & 3.20$^{***}$ (0.01) & 3.20$^{***}$ (0.01) & 2.23$^{***}$ (0.07) & 2.26$^{***}$ (0.07) \\ 
 \hline \\[-1.8ex] 
Observations & 5,991 & 5,971 & 5,889 & 5,882 \\ 
R$^{2}$ & 0.0004 & 0.001 & 0.14 & 0.15 \\ 
Adjusted R$^{2}$ & 0.0002 & 0.001 & 0.14 & 0.14 \\ 
Residual Std. Error & 0.74 (df = 5989) & 0.76 (df = 5969) & 0.68 (df = 5862) & 0.70 (df = 5855) \\ 
\hline 
\hline \\[-1.8ex] 
 & \multicolumn{4}{r}{$^{*}$p$<$0.1; $^{**}$p$<$0.05; $^{***}$p$<$0.01} \\ 
 & \multicolumn{4}{r}{Note: Columns 1,3 = worker voice treatment. Columns 2, 4 = financial incentives.} \\ 
\end{tabular} 
\end{table}

One possibility is that financial incentives for using the technology signal a company that is growing and willing to invest in its employees as it does so. An employer like this also might be more likely to value its employees (which is positively associated with the perceived impact of automation). It is also possible that workers see an employer's willingness to offer financial incentives as a sign of its financial health, making it less likely to lay workers off.

An alternative explanation is that the positive effect of financial incentives on perceived job security is that workers have self-belief that increased incentives will also improve their performance, which will translate into increased job security. Note that the financial incentives for workers after the introduction of Alpha are not guaranteed. They are only available to workers that make improvements in their productivity. However, workers receiving this treatment have optimistic responses on average about the downstream effects of the technology.

\begin{table}[!htb]
\begin{tabular}{|m{0.05\linewidth} | m{0.33\linewidth} | m{0.26\linewidth}  | m{0.25\linewidth} |}
 \hline
 Hyp & \makecell{Description} & \makecell{Measure} & \makecell{Result} \\ 
 \hline
H4A &
  \makecell{If workers are involved in the design\\ of automation, they will think more  \\positively about the impact of  \\automation.} &
  \makecell{Experimental design with \\ worker input treatment.} &
  \makecell{Hypothesis not supported. \\ 
  \textbf{Table 10, columns 1,3} \\ 
  Row: Treatment (Voice) \\} \\
\hline
H4B &
  \makecell{If workers have financial incentives to \\ embrace automation,\\ they will be more likely to do so.} &
  \makecell{Experimental design with \\ financial incentives treatment.} &
  \makecell{Hypothesis supported. \\ 
  \textbf{Table 10, columns 2,4} \\ 
  Row: Treatment (Bonus) \\ } \\
  \hline
\end{tabular}
\caption{Summary of experimental hypotheses and results}
\label{table:experimental}
\end{table}




\subsection{Technology appearance}

The hypotheses about technology design predicted that approaches to technology appearance would vary by a worker's physical environment (as a proxy for whether they had been exposed previously to similar technologies) and the environment in which the technology was presented.

We anticipated that workers in industries with more direct experience with automation would think differently about technology designs. However, workers in warehousing and production environments did not report significantly different perceptions of humanoid robots, robot arms, or mobility solutions than workers in other settings, aside from a few exceptions. This finding may be due to the complex suite of factors, e.g. autonomy, imitation, predictability, etc., beyond just appearance that dictate people's perception of technologies~\cite{zlotowski2020-one}. Results were likely affected as well by the small sample size for each subcategory. \footnote{The exceptions were that those who work in warehouse and production environments are more likely to respond that arms will be more helpful in a hospital setting. These workers are also more likely to report that the more playful-looking humanoid and robot arm are more dangerous than their machine-like counterparts in a construction setting. And finally, these workers in warehousing and construction settings are more likely to say that a conveyor belt is less dangerous than an autonomous mobile robot in a construction setting. There is no apparent pattern to the exceptions.} 

\begin{figure}
    \centering
    \includegraphics[width=.8\linewidth]{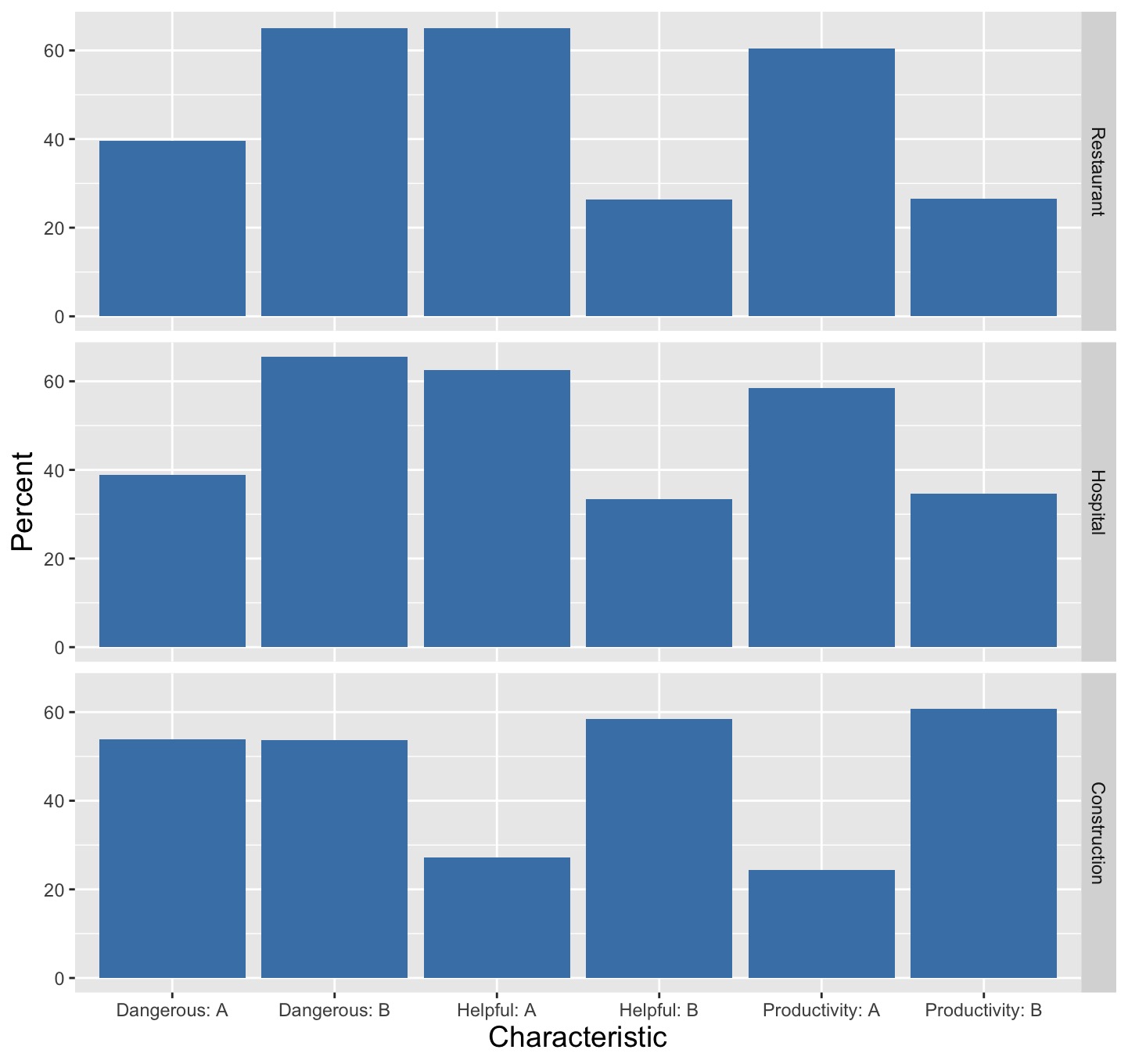}
    \caption{Perception of humanoid technologies by setting (Restaurant, Hospital, and Construction). Humanoid A has a more ``playful'' appearance, while humanoid B has a more ``machine-like'' appearance. }
    \label{fig:humanoids-settings}
\end{figure}

Workers' perspectives on different automation technologies varied by the settings in which the technologies were presented. Consistent with Hypothesis H5B and prior research, workers favored playful designs in less severe environments and more machine-like designs in more severe environments. The relationship is striking in the case of humanoid robots (see Figure \ref{fig:humanoids-settings}). Workers were asked to compare the potential danger, helpfulness, and productivity of the playful-looking humanoid (A) and the more machine-like humanoid (B). Respondents felt that the more playful-looking humanoid would be more helpful in a restaurant setting, whereas the more machine-like humanoid would be more dangerous. However, in a construction setting, participants were likelier to respond that the machine-like humanoid would be more helpful. These results suggest that workers believe the form of automation should match its function. \footnote{One result that contradicts Hypothesis 5B is that participants were more likely to indicate that the autonomous mobile robot would be helpful in the construction setting. This could be explained by prior knowledge of conveyor belts and the constantly-changing nature of construction, where the overhead necessary to set up infrastructure for conveyor belts could make them less practical.}

\begin{table}[!htb]
\begin{tabular}{|m{0.05\linewidth} | m{0.33\linewidth} | m{0.26\linewidth}  | m{0.25\linewidth} |}
 \hline
 Hyp & \makecell{Description} & \makecell{Measure} & \makecell{Result} \\
 \hline
H5A &
  \makecell{Workers with experience in\\ production environments will \\have different perceptions \\of technology impact.} &
  \makecell{Binomial logistic regression \\models with membership\\ in warehouse/fulfillment\\ group as independent variable.} &
  \makecell{No significant association. }\\
\hline
H5B &
  \makecell{Humans will respond most \\positively to machines that match \\in design the severity of\\ the task environment.} &
  \makecell{Head to head comparison \\of technologies with paired\\ Wilcoxon signed rank test} &
  \makecell{Hypothesis supported for \\most categories. } \\
  \hline
\end{tabular}
\caption{Summary of technology appearance hypotheses and results}
\label{table:appearance}
\end{table}

\section{Conclusion}

The impact of new technologies on work is often discussed as a story of winners and losers. The beneficiaries of automation are individuals with higher education in jobs that cannot be automated. The losers are those in routine jobs where technology can take over their work. This study deconstructs that narrative in several ways. 

First, it adds nuance to the common dichotomy of routine and non-routine jobs. Many workers identify that they frequently do routine tasks and cognitive tasks. Their jobs do not fit neatly into an ``automatable'' or ``safe from automation'' category. And when we examine the link between routine tasks and attitudes toward automation, we find that the connection is weak. Under some conditions routine tasks are associated with a more negative attitude toward automation, but the association is not always statistically significant. 

Primarily, it appears that one's concentration of cognitive tasks is the most reliable predictor of a positive attitude toward automation and its impact on work. One interpretation of this data is that if a worker is performing routine tasks, but also has problem-solving to do, they might be optimistic about the potential for technologies to help them manage their routine work and free them up to focus more on problem-solving. If the same worker did not have the problem-solving to do, they might be more resistant to using new technologies.

Second, the survey shows that after controlling for other factors, more education is not a strong predictor of a positive attitude toward automation in general --- or a perception that automation is having a positive impact on one's job security. If anything, it's the opposite. More education is associated with less optimism about the impact of new technologies and automation on one's job security. 

Consider multiple potential explanations for this phenomenon. One is that workers with less education are merely less aware of the threat that technologies pose. Their optimism in this assessment is unwarranted. However, this is inconsistent with other data points in the survey that indicate workers with more exposure to new technologies are also more optimistic about the impact of new technologies on their job security.

A second explanation is that workers are optimistic about the impact of automation on their jobs, but are pessimistic about the impact of automation overall. This perception is comparable to individuals approving of their Congressperson, but disapproving of Congress overall. The localized optimism about one's job does not mean that workers without a college degree overall broadly approve of automation. They could just be comfortable in their job. After all, U.S. job satisfaction in the survey is strikingly high, with 64\% of respondents signaling strong job satisfaction.

However, the data do not fully support this explanation. The survey asked workers a general question about the impact of automation, allowing them to select a number of potential effects of automation on the economy more broadly. About half of the workers selected ``increased layoffs'' as one of the effects of automation. Around the same percentage selected "decreased job security" as one of the effects. This is consistent with the net neutral impact that U.S. workers said automation would have on their jobs, suggesting that there is not a clear gap between a worker's perceived impact of automation on their own job versus impact on society in general.

A third explanation is that workers without a college degree, having seen technologies introduced to augment their work, realize that replacing workers with automation is really hard and therefore they are not worried. Or, consistent with studies of robot adoption that link technology adoption to more hiring, workers associate the introduction of new technologies with a growth outlook for a company, which makes them feel more secure in their jobs.

The third explanation is closest to the reality that we have seen in field research on the deployment of automation in factory and office environments. The deployment of new technologies is often messy, and its impact on work incremental. Even when companies try to reduce their workforce using technology, they have difficulty doing so.

One of the primary goals of this study is to shift the focus away from who wins and who loses from automation to how automation and new technologies can make work better -- according to workers. 

Four primary lessons from the survey stand out for employers.

\begin{enumerate}
    \item \textbf{Employees who like their jobs are more likely to embrace new technologies and automation.} Employees with high job satisfaction and who feel valued by their employer, as well as employees who say that their employer is invested in their upward mobility, all are associated with more optimistic attitudes toward the impact of automation on their work. For companies that are integrating new technologies and would like their workers to champion it, this core lesson suggests going back to basics: building trust with employees and ensuring they feel that their employer wants them to succeed.
    \item \textbf{Employees that see tangible benefits from automation also feel more secure about its impact.} The relatively benign suggestion that employees would be eligible for a bonus if they used a fictitious new technology productively led workers to become more optimistic about the impact of that technology on their work. It might seem obvious that workers respond to financial incentives, but in our field research, we have only rarely seen these incentives programs in practice. The key takeaway for employers is to help employees answer the question, ``What's in it for me?'' when they are introduced to new technology tools.
    \item \textbf{Employers can potentially design jobs in which workers are more supportive or more resistant to new technologies.} Several findings suggest that there are work environments in which workers have a more optimistic attitude toward the impact of automation on their jobs. Employers have control over some of these characteristics. For example, employers can design jobs to include frequent problem-solving alongside routine tasks to keep workers cognitively engaged and potentially more open to embracing technology for routine tasks. Moreover, employers can expose more of their employees to work with advanced technologies, particularly those who might show other characteristics of seeing benefits of new technologies for their work. 
    \item \textbf{Exposure to new technologies might also help employers identify new advocates for technology.} For example, Black and Hispanic workers without a college degree are among the most optimistic about the impact of technological change on their jobs, but they say they are less likely than their white peers to be a champion for new technology on the job. Only 33\% of Black workers without a college degree identify as a technology advocate within their organizations,, but 39\% of white workers do. This is in sharp contrast to college-educated workers of all races. A majority of white, Black, and Hispanic workers with a college degree say they champion new technologies within their organizations.
\end{enumerate}

A broader takeaway from the survey is about upward mobility and the desire to learn. Workers were asked how important it was to them to ``develop new skills and grow in their career.'' They also asked how much their employer invested in supporting their development of new skills and career growth. And finally, they evaluated how much they thought new technologies could contribute to their development of new skills and career growth. 

Their responses to these questions about upward mobility were revealing. For workers overall, upward mobility was not the highest priority. On average, developing new skills was not as important to them as safety and comfort at work or autonomy on the job. But when asked about how employers invest in their upward mobility, workers were more likely to say their employers ``never'' or ``rarely'' do so than they were to say that such investment is a frequent or constant occurrence. What makes this interesting is that workers in the United States were most optimistic about the potential for automation and new technologies to help them develop new skills and grow in their careers. And those workers for whom upward mobility is a priority were also more likely to see benefits from automation.

The story these data seem to tell is that there is untapped potential for new technologies and automation to help employers support workers as they develop new skills and grow in their careers. 

How can new technologies make jobs better? One indicator would be if they lead more employers to invest in skill development and upward mobility for workers, which could have potential spillover effects on workers' safety, comfort, pay, autonomy, and security on the job. 

\newpage

\bibliographystyle{ieeetr}
\bibliography{Automation_Worker_Perspective}

\newpage
\section{Appendix}

The Appendix includes supplementary materials that can add additional context to the robustness of the findings in the body of the paper. Table 13 is a glossary of the key variables in the paper, including how each was measured. Table 14 is a replication of Table 7 as a binomial logit model with "job security" as the binary outcome. The results are similar in both models. Tables 15 and 16 provide additional descriptive statistics of workers' characteristics and attitudes by race and education. They show that workers are in many ways similar across these categories on many variables, but there is a sharp divide when it comes to the perceived impact of automation. 

Table 17 replicates Table 10, except with job security as the binary outcome of a binomial logit model. It yields consistent results. 

Table 18 summarizes how workers responded to several questions about the technology designs they were assigned. Would they be interested in receiving training on the technology? Did they believe that this technology would replace jobs?

Although the sampling and framing of these questions is different than the survey's core questions about worker's perceptions of the impact of automation on their jobs (e.g. workers were randomly assigned to one technology category, so not every worker responds to every technology design), there are suggestive findings that emerge from workers' responses. The first is that workers are both eager to learn and concerned about the consequences of these technologies. This finding reinforces the descriptive statistics above. When asked about the impact of automation on their ability to develop new skills, workers were positive. When asked about the impact of automation on their job security, workers were neutral. Worker responses to the impact of these particular technologies capture this tension: automation represents a learning opportunity on the one hand and a potential source of displacement on the other.

The second finding is that when workers are asked about the likelihood of particular technologies to displace jobs, they offer net negative responses. However, when they are asked about the potential impact of automation on job displacement in their communities --- or on their own job security --- they are neutral. One potential explanation for this gap is that particular robotic technologies surface for workers the public discourse around automation displacing jobs. By agreeing to ``I believe this technology will displace jobs,'' workers are potentially signaling that they recognize and subscribe to the public narrative about robots and jobs.

Additional resources for the paper, including a summary of the questions asked and the crosstabular responses can be found online.

\begin{longtable}[c]{|l|l|}
\hline
Variable &
  Measure \\ \hline
\endfirsthead
\endhead
Age &
  Age in years at time of the survey. \\ \hline
Sex &
  Binary variable where female = 1. \\ \hline
Full-time &
  \begin{tabular}[c]{@{}l@{}}Binary variable where full-time is "Employed full-time (35 hours or more per week)\\ for pay with an organization or company.\end{tabular} \\ \hline
Trust &
  \begin{tabular}[c]{@{}l@{}}Index variable averaging likert responses (strongly disagree to strongly agree)\\ to four questions, including: "I usually trust people until they give me a reason not to \\ trust them."\end{tabular} \\ \hline
Education &
  \begin{tabular}[c]{@{}l@{}}Variable with four values (less than high school, high school, some college, Bachelors degree\\ or higher) roughly proportionate to years of education\end{tabular} \\ \hline
Income &
  Household income with 7 tranches from less than $10,000 to $150,000 or more. \\ \hline
Physical jobs &
  \begin{tabular}[c]{@{}l@{}}Binary variable in which workers in operations, production, and warehousing jobs are 1. \\ Alternative jobs are listed as Office jobs in multiple tables.\end{tabular} \\ \hline
Routine tasks &
  \begin{tabular}[c]{@{}l@{}}Binary variable in which workers who list their frequency of routine tasks as \\ "Most of my time" or "All of my time" are 1.\end{tabular} \\ \hline
Problem solving &
  \begin{tabular}[c]{@{}l@{}}Binary variable in which workers who list their frequency of solving complex problems as \\ "Most of my time" or "All of my time" are 1.\end{tabular} \\ \hline
Job satisfaction &
  \begin{tabular}[c]{@{}l@{}}Index variable averaging workers' satisfaction with various aspects of their job, \\ including their pay, productivity, upward mobility, autonomy, safety and comfort, \\ and job security.\end{tabular} \\ \hline
\begin{tabular}[c]{@{}l@{}}Valued by \\ employer\end{tabular} &
  \begin{tabular}[c]{@{}l@{}}Binary variable in which workers who say they are very valued or extremely valued by their \\ employer are 1.\end{tabular} \\ \hline
Tech adoption &
  \begin{tabular}[c]{@{}l@{}}Index variable averaging how often workers interact with four technologies: \\ software automation, robotics, artificial intelligence, and generative artificial intelligence. \\ Higher values indicate more frequent uses of advanced technology.\end{tabular} \\ \hline
Video games &
  \begin{tabular}[c]{@{}l@{}}Incremental variable indicating how often workers played video games in the last month \\ with each increment representing five times.\end{tabular} \\ \hline
Screen time &
  \begin{tabular}[c]{@{}l@{}}Incremental variable indicating how often workers spent on screens during a typical day \\ they are not working with increments ranging from less than one hour to more than 8 hours.\end{tabular} \\ \hline
Early adopter &
  \begin{tabular}[c]{@{}l@{}}Binary variable with workers who are "among the first people" to use new technology, or \\ "sooner than most people" are 1.\end{tabular} \\ \hline
Champion &
  \begin{tabular}[c]{@{}l@{}}Binary variable with workers who agree that "I am a strong advocate for the usage of the\\ technology in my workplace" as 1.\end{tabular} \\ \hline
Troubleshooter &
  \begin{tabular}[c]{@{}l@{}}Binary variable with workers who say they "Troubleshoot technology to solve problems \\ that arise during use" very often or all of the time as 1.\end{tabular} \\ \hline
Learner &
  \begin{tabular}[c]{@{}l@{}}Binary variable where workers who say it's very important or extremely important for \\ them to know "how the technology tool works (what's going on under the hood)" for them\\ to do their job well are 1.\end{tabular} \\ \hline
Empl inv prod. &
  \begin{tabular}[c]{@{}l@{}}Binary variable where workers say their employer invests frequently or all the time in \\ "improving your productivity" are 1.\end{tabular} \\ \hline
Empl inv pay &
  \begin{tabular}[c]{@{}l@{}}Binary variable where workers say their employer invests frequently or all the time in \\ "bonuses or financial incentives for good job performance that boost your pay" are 1.\end{tabular} \\ \hline
\begin{tabular}[c]{@{}l@{}}Empl inv \\ ergonomics\end{tabular} &
  \begin{tabular}[c]{@{}l@{}}Binary variable where workers say their employer invests frequently or all the time \\ "in the safety and comfort of how you do your job" are 1.\end{tabular} \\ \hline
\begin{tabular}[c]{@{}l@{}}Empl inv \\ autonomy\end{tabular} &
  \begin{tabular}[c]{@{}l@{}}Binary variable where workers say their employer invests frequently or all the time in giving\\ "you additional control over what you do and when you do it at work" are 1.\end{tabular} \\ \hline
Emp inv mobility &
  \begin{tabular}[c]{@{}l@{}}Binary variable where workers say their employer invests frequently or all the time in \\ "your ability to develop new skills and grow in your career" are 1.\end{tabular} \\ \hline
Imp. of pay &
  Binary variable where workers say "your pay" is very important or extremely important are 1. \\ \hline
\begin{tabular}[c]{@{}l@{}}Imp. of \\ ergonomics\end{tabular} &
  \begin{tabular}[c]{@{}l@{}}Binary variable where workers say "they safety or comfort of how you do your job" is very \\ important or extremely important are 1.\end{tabular} \\ \hline
Imp. of autonomy &
  \begin{tabular}[c]{@{}l@{}}Binary variable where workers say "the amount of control over what you do and when you \\ do it at work" is very important or extremely important are 1.\end{tabular} \\ \hline
Imp. of mobility &
  \begin{tabular}[c]{@{}l@{}}Binary variable where workers say "your ability to develop new skills and grow in \\ your career" is very important or extremely important are 1.\end{tabular} \\ \hline
\begin{tabular}[c]{@{}l@{}}Coord. Market \\ Econ.\end{tabular} &
  Binary variable where workers from Germany, France, Spain, and Italy are 1. \\ \hline
\begin{tabular}[c]{@{}l@{}}Lib. Market \\ Econ.\end{tabular} &
  \begin{tabular}[c]{@{}l@{}}Binary variable where workers from the United States, Australia, \\ and the United Kingdom are 1.\end{tabular} \\ \hline
\caption{Glossary of variables in regression models}
\label{tab:my-table}\\
\end{longtable}

\begin{table}[!htbp] \centering 
  \caption{Predictors of Perceived Impact of Automation on Job Security} 
  \label{} 
\footnotesize 
\begin{tabular}{@{\extracolsep{5pt}}lccccc} 
\\[-1.8ex]\hline 
\hline \\[-1.8ex] 
 & \multicolumn{5}{c}{\textit{Dependent variable:}} \\ 
\cline{2-6} 
\\[-1.8ex] & \multicolumn{5}{c}{Positive impact of automation on own job security} \\ 
\\[-1.8ex] & (1) & (2) & (3) & (4) & (5)\\ 
\hline \\[-1.8ex] 
 Age & 0.004 (0.002) & 0.001 (0.003) & 0.004 (0.003) & 0.003 (0.002) & 0.003 (0.002) \\ 
  Sex (Female) & 0.11$^{*}$ (0.07) & 0.19$^{**}$ (0.07) & 0.19$^{**}$ (0.07) & 0.14$^{***}$ (0.05) & 0.14$^{***}$ (0.05) \\ 
  Full-time & 0.28$^{***}$ (0.08) & 0.22$^{**}$ (0.09) & 0.16$^{*}$ (0.09) & 0.07 (0.06) & 0.05 (0.06) \\ 
  Trust & 0.18$^{***}$ (0.04) & 0.09$^{**}$ (0.04) & 0.07$^{*}$ (0.04) & 0.10$^{***}$ (0.03) & 0.10$^{***}$ (0.03) \\ 
  Education & $-$0.19$^{***}$ (0.04) & $-$0.29$^{***}$ (0.04) & $-$0.29$^{***}$ (0.04) &  &  \\ 
  Income & $-$0.01 (0.02) & $-$0.04 (0.03) & $-$0.03 (0.03) &  &  \\ 
  Physical jobs & 0.17$^{**}$ (0.08) & 0.18$^{**}$ (0.09) & 0.18$^{**}$ (0.09) & 0.25$^{***}$ (0.06) & 0.25$^{***}$ (0.06) \\ 
  Routine tasks & 0.02 (0.07) & 0.11 (0.07) & 0.07 (0.07) & 0.19$^{***}$ (0.05) & 0.19$^{***}$ (0.05) \\ 
  Problem solving & 0.65$^{***}$ (0.07) & 0.36$^{***}$ (0.08) & 0.27$^{***}$ (0.08) & 0.14$^{**}$ (0.06) & 0.14$^{**}$ (0.06) \\ 
  Job satisfaction &  & 0.24$^{***}$ (0.05) & 0.18$^{***}$ (0.05) & 0.14$^{***}$ (0.04) & 0.13$^{***}$ (0.04) \\ 
  Valued by employer &  & 0.33$^{***}$ (0.08) & 0.29$^{***}$ (0.09) & 0.17$^{***}$ (0.06) & 0.17$^{***}$ (0.06) \\ 
  Tech adoption &  & 0.44$^{***}$ (0.05) & 0.42$^{***}$ (0.05) & 0.42$^{***}$ (0.03) & 0.43$^{***}$ (0.03) \\ 
  Video games &  & $-$0.13$^{***}$ (0.02) & $-$0.13$^{***}$ (0.02) & $-$0.07$^{***}$ (0.02) & $-$0.07$^{***}$ (0.02) \\ 
  Screen time &  & 0.03 (0.03) & 0.03 (0.03) & 0.06$^{**}$ (0.02) & 0.05$^{**}$ (0.02) \\ 
  Early adopter &  & 0.25$^{***}$ (0.08) & 0.24$^{***}$ (0.08) & 0.20$^{***}$ (0.06) & 0.20$^{***}$ (0.06) \\ 
  Champion &  & 0.54$^{***}$ (0.07) & 0.51$^{***}$ (0.08) & 0.47$^{***}$ (0.05) & 0.47$^{***}$ (0.05) \\ 
  Troubleshooter &  & $-$0.10 (0.09) & $-$0.10 (0.09) & 0.03 (0.06) & 0.03 (0.06) \\ 
  Learner &  & 0.47$^{***}$ (0.08) & 0.40$^{***}$ (0.08) & 0.22$^{***}$ (0.06) & 0.23$^{***}$ (0.06) \\ 
  Empl inv prod. &  &  & 0.13 (0.09) & 0.29$^{***}$ (0.07) & 0.30$^{***}$ (0.07) \\ 
  Empl inv pay &  &  & 0.06 (0.09) & 0.12$^{*}$ (0.07) & 0.11$^{*}$ (0.07) \\ 
  Emp inv ergonomics &  &  & 0.09 (0.09) & 0.13$^{**}$ (0.06) & 0.14$^{**}$ (0.06) \\ 
  Emp inv autonomy &  &  & $-$0.14$^{*}$ (0.08) & $-$0.19$^{***}$ (0.06) & $-$0.20$^{***}$ (0.06) \\ 
  Emp inv mobility &  &  & 0.12 (0.09) & 0.13$^{**}$ (0.07) & 0.13$^{**}$ (0.07) \\ 
  Importance of pay &  &  & 0.21$^{**}$ (0.09) & 0.11$^{*}$ (0.06) & 0.11$^{*}$ (0.06) \\ 
  Importance of ergonomics &  &  & 0.01 (0.08) & 0.02 (0.06) & 0.03 (0.06) \\ 
  Importance of autonomy &  &  & 0.04 (0.08) & $-$0.002 (0.06) & $-$0.002 (0.06) \\ 
  Importance of mobility &  &  & 0.27$^{***}$ (0.08) & 0.21$^{***}$ (0.06) & 0.21$^{***}$ (0.06) \\ 
  Coord. Market Econ. &  &  &  & 0.09 (0.06) &  \\ 
  Lib. Market Econ. &  &  &  &  & $-$0.21$^{**}$ (0.08) \\ 
  Constant & $-$1.64$^{***}$ (0.22) & $-$2.61$^{***}$ (0.29) & $-$2.65$^{***}$ (0.30) & $-$3.51$^{***}$ (0.20) & $-$3.46$^{***}$ (0.20) \\ 
 \hline \\[-1.8ex] 
Observations & 4,943 & 4,873 & 4,831 & 8,866 & 8,866 \\ 
Log Likelihood & $-$2,939.84 & $-$2,695.16 & $-$2,650.06 & $-$5,273.86 & $-$5,271.03 \\ 
Akaike Inf. Crit. & 5,899.68 & 5,428.32 & 5,356.12 & 10,601.71 & 10,596.05 \\ 
\hline 
\hline \\[-1.8ex] 
& \multicolumn{5}{r}{$^{*}$p$<$0.1; $^{**}$p$<$0.05; $^{***}$p$<$0.01} \\ 
 & \multicolumn{5}{r}{Note:Columns 1-3 are U.S.-specific models. Columns 4-5 are global models.} \\ 
\end{tabular} 
\end{table} 

\begin{table}[]
\centering
\resizebox{\columnwidth}{!}{
\begin{tabular}{llllllllll}
\toprule
Race & Education & Workers & Avg Age & Office jobs & Valued by employer & Routine & Problem-solving & Job satisfaction & Early adopter \\ \hline \\ 
Black    & no college & 263  & 45 & 46\%& 39\%& 49\%& 28\%& 70\%& 75\%\\
Hispanic & no college & 373  & 44 & 43\%& 43\%& 49\%& 25\%& 65\%& 69\%\\
Other    & no college & 113  & 45 & 50\%& 35\%& 49\%& 32\%& 52\%& 70\%\\
White    & no college & 1531 & 50 & 47\%& 51\%& 63\%& 23\%& 61\%& 71\%\\
Black    & college    & 184  & 48 & 55\%& 48\%& 60\%& 36\%& 66\%& 76\%\\
Hispanic & college    & 201  & 43 & 38\%& 49\%& 64\%& 38\%& 61\%& 75\%\\
Other    & college    & 257  & 46 & 40\%& 46\%& 58\%& 35\%& 60\%& 70\%\\
White    & college    & 1935 & 47 & 41\%& 55\%& 71\%& 34\%& 58\%& 67\%\\ \hline
\end{tabular}
}
\caption{Worker characteristics by race and education}
\label{tab:my-table}
\end{table}

\begin{table}[]
\centering
\resizebox{\columnwidth}{!}{
\begin{tabular}{lllllllllll}
\toprule
     &           &         & \multicolumn{4}{l}{Factors important to enjoying your work} & \multicolumn{4}{l}{Factors your employer is invested in} \\ \hline \\ 
Race & Education & Workers & Safety \& comfort   & Pay  & Autonomy  & Upward mobility  & Safety \& comfort  & Pay  & Autonomy  & Upward mobility  \\ \hline \\
Black    & no college & 263  & 53\%& 52\%& 47\%& 24\%& 37\%& 35\%& 51\%& 36\%\\
Hispanic & no college & 373  & 53\%& 41\%& 35\%& 19\%& 40\%& 24\%& 47\%& 40\%\\
Other    & no college & 113  & 47\%& 34\%& 32\%& 17\%& 40\%& 25\%& 41\%& 31\%\\
White    & no college & 1531 & 50\%& 33\%& 34\%& 18\%& 43\%& 23\%& 38\%& 22\%\\
Black    & college    & 184  & 63\%& 52\%& 41\%& 22\%& 48\%& 34\%& 54\%& 35\%\\
Hispanic & college    & 201  & 55\%& 55\%& 40\%& 25\%& 50\%& 39\%& 47\%& 34\%\\
Other    & college    & 257  & 61\%& 42\%& 37\%& 23\%& 48\%& 29\%& 46\%& 27\%\\
White    & college    & 1935 & 62\%& 41\%& 31\%& 20\%& 48\%& 32\%& 40\%& 21\%\\ \hline
\end{tabular}
}
\caption{Job attitudes by race and education}
\label{tab:my-table}
\end{table}

\begin{table}[!htbp] \centering 
  \caption{Experimental Results (Job Security)} 
  \label{} 
\footnotesize 
\begin{tabular}{@{\extracolsep{0pt}}lcccc} 
\\[-1.8ex]\hline 
\hline \\[-1.8ex] 
 & \multicolumn{4}{c}{\textit{Dependent variable:}} \\ 
\cline{2-5} 
\\[-1.8ex] & \multicolumn{4}{c}{Positive impact of automation on own job security} \\ 
\\[-1.8ex] & (1) & (2) & (3) & (4)\\ 
\hline \\[-1.8ex] 
 Treatment (Voice) & $-$0.07 (0.06) &  & $-$0.04 (0.06) &  \\ 
  Treatment (Bonus) &  & 0.12$^{**}$ (0.06) &  & 0.15$^{**}$ (0.06) \\ 
  Age &  &  & 0.003 (0.003) & 0.003 (0.003) \\ 
  Sex (Female) &  &  & 0.03 (0.07) & 0.05 (0.07) \\ 
  Full-time &  &  & 0.05 (0.08) & 0.14$^{*}$ (0.08) \\ 
  Trust &  &  & 0.01 (0.04) & 0.07$^{**}$ (0.04) \\ 
  Physical jobs &  &  & 0.12 (0.08) & 0.13$^{*}$ (0.08) \\ 
  Routine tasks &  &  & 0.22$^{***}$ (0.07) & 0.14$^{**}$ (0.06) \\ 
  Problem solving &  &  & 0.05 (0.07) & $-$0.03 (0.07) \\ 
  Job satisfaction &  &  & 0.07 (0.05) & 0.01 (0.05) \\ 
  Valued by employer &  &  & 0.08 (0.08) & 0.20$^{***}$ (0.08) \\ 
  Tech adoption &  &  & 0.52$^{***}$ (0.04) & 0.49$^{***}$ (0.04) \\ 
  Video games &  &  & $-$0.05$^{**}$ (0.02) & $-$0.03 (0.02) \\ 
  Screen time &  &  & 0.05 (0.03) & 0.07$^{**}$ (0.03) \\ 
  Early adopter &  &  & 0.26$^{***}$ (0.07) & 0.07 (0.07) \\ 
  Champion &  &  & 0.46$^{***}$ (0.07) & 0.45$^{***}$ (0.07) \\ 
  Troubleshooter &  &  & 0.01 (0.08) & 0.04 (0.08) \\ 
  Learner &  &  & 0.27$^{***}$ (0.07) & 0.21$^{***}$ (0.07) \\ 
  Empl inv prod. &  &  & 0.16$^{*}$ (0.09) & 0.22$^{***}$ (0.08) \\ 
  Empl inv pay &  &  & 0.15$^{*}$ (0.08) & 0.23$^{***}$ (0.08) \\ 
  Emp inv ergonomics &  &  & 0.34$^{***}$ (0.08) & 0.12 (0.08) \\ 
  Emp inv autonomy &  &  & $-$0.23$^{***}$ (0.08) & $-$0.13$^{*}$ (0.07) \\ 
  Emp inv mobility &  &  & 0.13 (0.09) & 0.06 (0.08) \\ 
  Importance of pay &  &  & 0.01 (0.08) & 0.19$^{**}$ (0.08) \\ 
  Importance of ergonomics &  &  & $-$0.05 (0.08) & $-$0.07 (0.08) \\ 
  Importance of autonomy &  &  & $-$0.15$^{*}$ (0.08) & $-$0.13$^{*}$ (0.07) \\ 
  Importance of mobility &  &  & 0.17$^{**}$ (0.08) & 0.32$^{***}$ (0.07) \\ 
  Constant & $-$0.99$^{***}$ (0.04) & $-$0.99$^{***}$ (0.04) & $-$3.15$^{***}$ (0.26) & $-$3.36$^{***}$ (0.25) \\ 
 \hline \\[-1.8ex] 
Observations & 5,983 & 5,969 & 5,882 & 5,881 \\ 
Log Likelihood & $-$3,708.48 & $-$3,841.18 & $-$3,264.70 & $-$3,388.48 \\ 
Akaike Inf. Crit. & 7,420.96 & 7,686.37 & 6,583.40 & 6,830.96 \\ 
\hline 
\hline \\[-1.8ex] 
\textit{Note:}  & \multicolumn{4}{r}{$^{*}$p$<$0.1; $^{**}$p$<$0.05; $^{***}$p$<$0.01} \\ 
 & \multicolumn{4}{r}{Note:Columns 1 and 3 are the worker voice treatment. Columns 2 and 4 are the financial incentives treatment.} \\ 
\end{tabular} 
\end{table} 

\begin{table}[]
\centering
\begin{tabular}{lllll}
\toprule
 &
  \multicolumn{2}{l}{\begin{tabular}[c]{@{}l@{}}I would be interested in receiving training \\ to operate this technology\end{tabular}} &
  \multicolumn{2}{l}{\begin{tabular}[c]{@{}l@{}}I believe this technology\\ will displace jobs\end{tabular}} \\
  \hline \\
\emph{Technology} & \emph{Agree}  & \emph{Disagree} & \emph{Agree}  & \emph{Disagree} \\
Pepper     & 47.0\% & 28.4\%   & 56.9\% & 17.8\%   \\
Phoenix    & 44.2\% & 32.2\%   & 58.1\% & 16.1\%   \\
Dobot      & 46.5\% & 26.1\%   & 46.5\% & 17.8\%   \\
DZGH       & 43.9\% & 28.7\%   & 51.6\% & 16.1\%   \\
AMR        & 37.7\% & 29.5\%   & 37.7\% & 29.5\%   \\
Conveyor   & 47.1\% & 24.7\%   & 42.7\% & 20.2\%   \\
\hline \\
\end{tabular}

\caption{Perceptions of automation technology by design}
\label{tab:design-perceptions}
\end{table}

\newpage

\begin{figure}
    \centering
    \includegraphics[width=1\linewidth]{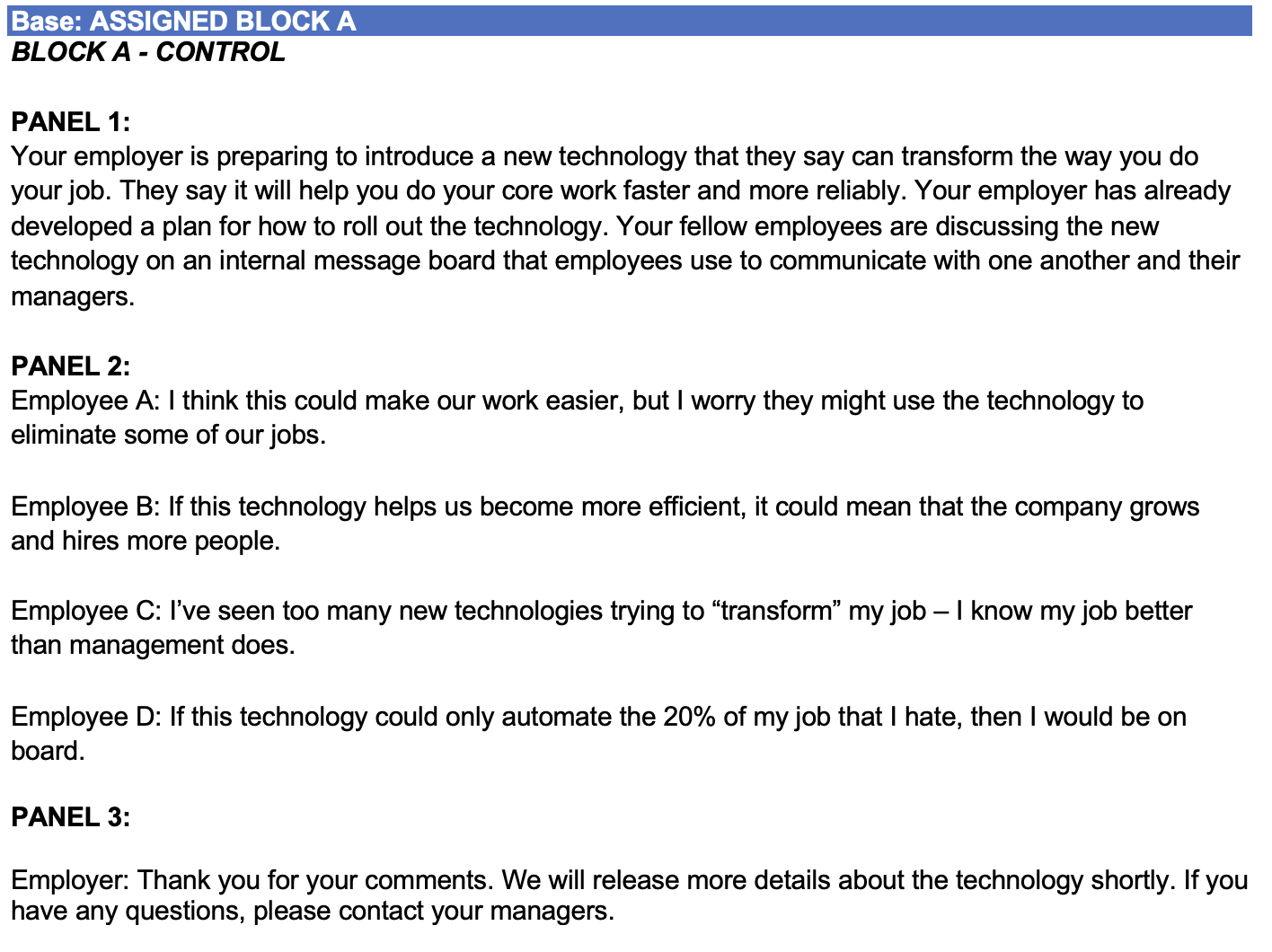}
    \caption{Control Block for Experiment}
    \label{fig:enter-label}
\end{figure}

\begin{figure}
    \centering
    \includegraphics[width=1\linewidth]{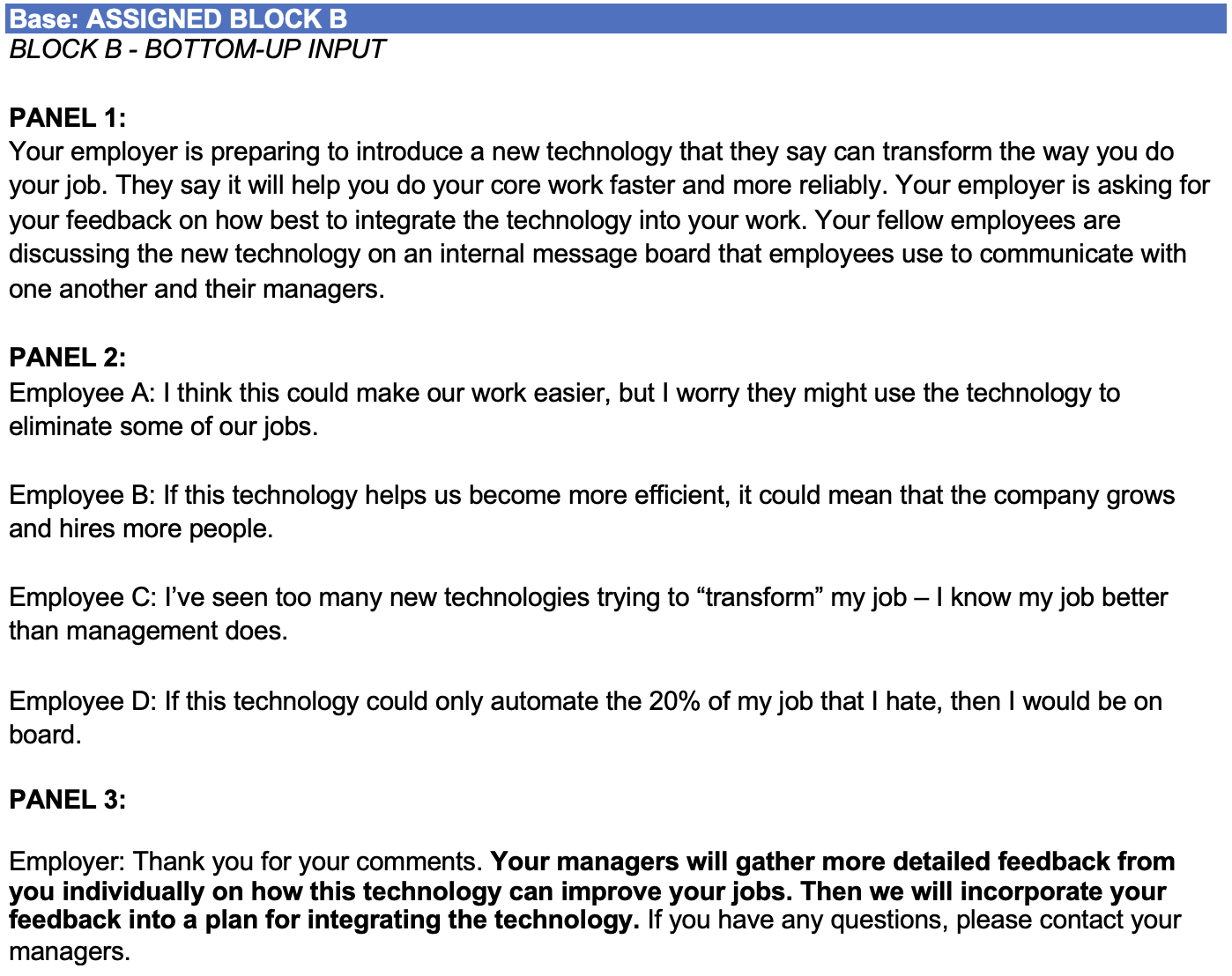}
    \caption{Treatment (Voice) Experimental Block}
    \label{fig:enter-label}
\end{figure}

\begin{figure}
    \centering
    \includegraphics[width=1\linewidth]{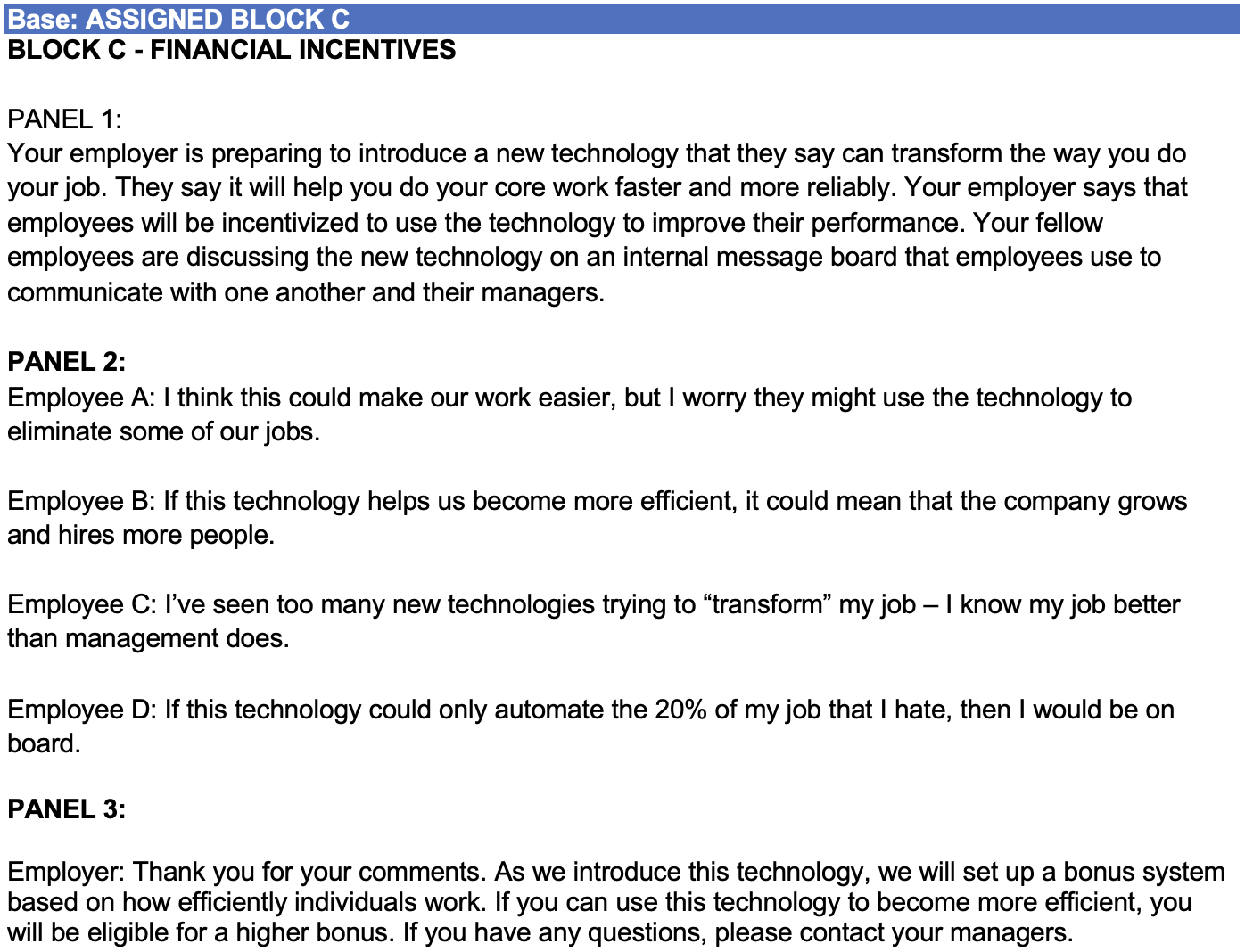}
    \caption{Treatment (Bonus) Experimental Block}
    \label{fig:enter-label}
\end{figure}

\end{document}